\documentclass[a4paper, 11pt]{article}
\usepackage[T1]{fontenc}
\usepackage[utf8]{inputenc}
\usepackage[english]{babel}
\usepackage{amsmath}
\usepackage{amssymb}
\usepackage{braket}
\usepackage{mathtools}
\usepackage{dsfont}
\usepackage{subcaption}
\usepackage{array}
\usepackage{booktabs}
\usepackage{yfonts}
\usepackage{setspace}
\usepackage{verbatim}

\usepackage[usenames,dvipsnames]{color}

\numberwithin{equation}{section}

\def\be{\begin{equation}}
\def\ee{\end{equation}}

\newcommand{\diff}{\mathrm{d}}
\def\aa{\mathtt{a}}
\def\cc{\mathtt{c}}
 
\definecolor{applegreen}{rgb}{0.55, 0.71, 0.0}

\usepackage{jheppub}                   % jheppub.sty

\makeatletter
\gdef\@fpheader{\ }                    % hack the jhep header
\makeatother

\title{Boundary terms and conserved charges in higher-derivative gauged supergravity}

\author[a]
{Davide Cassani,}
\author[c]
{Alejandro Ruip\'erez,}
\author[a,b]
{Enrico Turetta,}
\emailAdd{davide.cassani@pd.infn.it, alejandro.ruiperez@roma2.infn.it, enrico.turetta@phd.unipd.it}

\affiliation[a]{INFN, Sezione di Padova, Via Marzolo 8, 35131 Padova, Italy}
\affiliation[b]{Dipartimento di Fisica e Astronomia ``Galileo Galilei'', Universit\`a di Padova,\\Via Marzolo 8, 35131 Padova, Italy}
\affiliation[c]{Dipartimento di Fisica, Universit\`a di Roma ``Tor Vergata'' \& Sezione INFN Roma 2,\\ Via della Ricerca Scientifica 1, 00133, Roma, Italy}

\abstract{We address some issues in higher-derivative gauged supergravity with Chern-Simons terms, focusing on the five-dimensional case. We discuss the variational problem with Dirichlet boundary conditions as well as holographic renormalization in asymptotically locally AdS spacetimes, and derive the corresponding boundary terms. We then employ Wald's formalism in order to define conserved charges associated to local symmetries (diffeomorphisms and U(1) gauge transformations), taking into account the effect of generic gauge Chern-Simons terms. We prove that the first law of black hole mechanics and the quantum statistical relation hold in this setup. Chern-Simons terms also lead us to distinguish between Noether charges and Page (or Komar) charges which satisfy the Gauss law. We make use of the latter to compute corrections to the angular momentum and electric charge of the supersymmetric black hole in AdS$_5$ from its corrected near-horizon geometry. This also allows us to derive the microcanonical form of the entropy as a function of the conserved charges relying entirely on the near-horizon geometry. Finally, we comment on four-derivative gauged supergravity in four dimensions, showing that field redefinitions permit to simplify the action at linear order in the corrections, so that the equations of motion are those of the two-derivative theory.
}

\begin{document}

\maketitle

%%%%%%%%%%%%%%%%%%%%%%%%%%%%%%%%%%%%%%%%%%%%%%%%%%%%%%%%%%%%%%%%%%%%%%%%%%%%%%%%%%%%%%
\section{Introduction and summary of results}\label{sec:structure}

String theory is expected to provide a well-behaved infinite series of higher-derivative corrections to two-derivative supergravity in ten dimensions, controlled by the string tension~$\alpha'$. However this is only partially known, and its reduction to lower dimension crucially depends on the details of the compactification, besides mixing with quantum effects in the Kaluza-Klein towers. 
Given these limitations, a more agnostic way to proceed is to adopt an effective field theory approach where all higher-derivative terms compatible with the symmetries of interest are included as corrections to the lower-dimensional two-derivative theory. Studying the self-consistency and consequences of the corrections, one can aim at extracting the imprint of a high-energy completion. 
  Moreover, the perturbative treatment inherent to the effective field theory approach has the benefit of avoiding the issue that higher-derivative terms generically propagate extra degrees of freedom, including negative energy excitations~\cite{Stelle:1977ry}. 

Black hole thermodynamics still makes sense in the presence of higher-derivative corrections: for a generic diffeomorphism invariant higher-derivative action, the corrections to the two-derivative Bekenstein-Hawking area law for the entropy are computed by Wald's formula \cite{Wald:1993nt,Iyer:1994ys}. Moreover, for extremal black holes Sen's formalism based on the near-horizon geometry provides a convenient method to evaluate the Wald entropy and opens the way for the definition of the full quantum entropy \cite{Sen:2005wa,Sen:2007qy}.

From a holographic perspective, higher-derivative corrections in asymptotically locally AdS (AlAdS) spacetimes correspond to subleading terms in a large-$N$ expansion of the dual conformal field theory. 
When some supersymmetry is preserved, both the gravity and the field theory sides of the duality are better under control, and it becomes possible to obtain a  precise match of many quantities, such as partition functions, conserved charges and higher-point correlation functions.

In the recent years, there has been significant progress towards reaching a detailed understanding of the thermodynamics of supersymmetric AdS black holes in different spacetime dimensions via holography. A microstate counting  based on superconformal field theory (SCFT) partition functions at large $N$ has been shown to reproduce precisely the Bekenstein-Hawking entropy of such black holes, starting with \cite{Benini:2015eyy,Benini:2016rke} in four dimensions and \cite{Cabo-Bizet:2018ehj,Choi:2018hmj,Benini:2018ywd} in five dimensions (see \cite{Zaffaroni:2019dhb} for a review). These field theory partition functions can also be studied beyond the strict large-$N$ limit, with the subleading terms in the large-$N$ expansion predicting specific higher-derivative contributions on the gravity side. For instance, the four-dimensional SCFT partition function providing the entropy of the five-dimensional BPS black hole of~\cite{Gutowski:2004ez,Chong:2005hr} can be studied at finite $N$ in a large-charge Cardy-like regime. One finds that in this regime the partition function (and thus the entropy) is controlled to a large extent by the superconformal anomaly coefficients of the theory \cite{Kim:2019yrz,Cabo-Bizet:2019osg,Lezcano:2021qbj,Amariti:2021ubd,Cassani:2021fyv,ArabiArdehali:2021nsx}.
Recently,  these subleading corrections in the large-$N$ expansion have been matched by evaluating the black hole action in higher-derivative five-dimensional supergravity~\cite{Bobev:2022bjm,Cassani:2022lrk}. There is also considerable progress in obtaining the subleading corrections to three-dimensional large-$N$ SCFT partition functions with a holographic dual, see e.g.~\cite{Hristov:2022lcw,Bobev:2022eus} for a sample of very recent results.

This motivates investigating further higher-derivative corrections to gauged supergravities admitting AdS solutions. Here, the presence of a dimensionful coupling constant (related to the AdS radius) makes the classification of  corrections more subtle than the one based on a simple derivative counting. One way to determine the higher-derivative supergravity action is to rely on off-shell methods to construct supersymmetric invariants, then integrate out the auxiliary fields, and finally implement suitable field redefinitions to simplify the final expression. This task is greatly simplified when one adopts the effective field theory approach and works perturbatively in the higher-derivative couplings.

In this paper we clarify some issues that arise when dealing with
higher-derivative gauged supergravity in four and five dimensions in the context outlined above. 
Relevant previous work in five dimensions appeared e.g.\ in~\cite{Hanaki:2006pj,Cremonini:2008tw,Baggio:2014hua,Bobev:2021qxx,Liu:2022sew,Bobev:2022bjm,Cassani:2022lrk}, while progress in four dimensions has been reported e.g.\ in~\cite{Bobev:2020egg,Bobev:2020zov,Bobev:2021oku,Genolini:2021urf}.  

In five dimensions, the minimal supergravity theory involves an Abelian gauge field $A$ in addition to the metric and the gravitino, and the two-derivative action contains a Chern-Simons term $A\wedge F \wedge F$, where $F = \diff A$. At the four derivative level, one should also include the Chern-Simons term $A\wedge R^{ab} \wedge R_{ab}$, where $R^{ab}$ is the Riemann curvature two-form, together with its supersymmetric completion constructed in \cite{Hanaki:2006pj,Ozkan:2013nwa}. These terms encode the R-current anomaly of a generic dual SCFT, and as such play a crucial role in the holographic match established in~\cite{Bobev:2022bjm,Cassani:2022lrk}.
In this paper we will refer to the bosonic action given in~\cite{Cassani:2022lrk}, which completes and simplifies previous expressions obtained in \cite{Hanaki:2006pj, Cremonini:2008tw,Baggio:2014hua, Bobev:2021qxx,Liu:2022sew} thanks to the use of convenient field redefinitions.

The first question we address concerns the boundary terms that need to be added to the five-dimensional gauged supergravity action. These are made by the generalized Gibbons-Hawking-York terms necessary to have a well-posed variational principle defining the equations of motion, and by the counterterms that remove the divergences appearing when the action is evaluated on an AlAdS solution. Generically these terms contain a finite piece and contribute to the final result for the on-shell action; it is thus important to take them into account when comparing the latter with a given field theory partition function. 

We start by discussing the variational principle for the action, and show that despite the higher-derivative terms, it makes sense to just impose Dirichlet boundary conditions in an AlAdS spacetime.
Although our focus is on the supergravity action given in~\cite{Cassani:2022lrk}, we also make some more general considerations. In particular, we identify a four-derivative coupling of the metric and gauge field curvatures that yields two-derivative equations and thus admits a good Dirichlet variational principle, independently of the asymptotics of the spacetime and of whether it is regarded as a small correction to a two-derivative action.
We next implement holographic renormalization using the Hamilton-Jacobi method; this generalizes the work of \cite{Fukuma:2001uf,Liu:2008zf} as we include a gauge field, and of~\cite{Landsteiner:2011iq,Cremonini:2009ih} as we add the  (bosonic) supersymmetric completion of the higher-derivative terms considered there. 
The outcome of our analysis confirms that the boundary terms introduced in \cite{Cassani:2022lrk} based on the study of the black hole solution 
 and used to match the SCFT partition function are indeed correct.

The second part of the paper discusses conserved charges and black hole thermodynamics in presence of higher-derivative and Chern-Simons terms. First we revisit Wald's formalism~\cite{Wald:1993nt,Iyer:1994ys} to define the conserved charges associated to diffeomorphisms and U(1) gauge transformations, carefully keeping track of the contribution from gauge Chern-Simons terms.\footnote{The effect of purely gravitational Chern-Simons terms (which we are not considering) has been discussed before in e.g.~\cite{Tachikawa:2006sz, Elgood:2020nls}.} We emphasize that it is important to distinguish between the charges defined through the Noether procedure and those which satisfy the Gauss law, which are often referred to as Komar or Page charges. We provide explicit expressions for the latter. In particular, we present the formulae for angular momenta and electric charge associated with our five-dimensional higher-derivative supergravity action.
We then make use of these general results to demonstrate the first law of black hole mechanics as well as to prove the quantum statistical relation which identifies the Euclidean on-shell action (with Dirichlet boundary conditions) with the Legendre transform of the black hole entropy. This is done working in a gauge that is regular up to the horizon, so as to be able to use the Stokes theorem to relate integrals  at the horizon and at the conformal boundary.
 
Our main motivation to work with charges that obey a Gauss law is that we can exploit them to compute the corrections to the angular momentum and the electric charge of the  supersymmetric AdS$_5$ black hole whose two-derivative solution was given in~\cite{Gutowski:2004ez}: although the full corrected solution is not known and thus we cannot evaluate the charges by means of an asymptotic integral at the boundary, we can obtain them from our new formulae applied to the corrected near-horizon solution found in~\cite{Cassani:2022lrk}. The results are in agreement with the expressions we derived in~\cite{Cassani:2022lrk} by varying the on-shell action, modulo a constant shift in the electric charge that we discuss. This computation also allows us to express the  black hole entropy as a function of the angular momentum and electric charge relying solely on the near-horizon solution.
We leave for future work a comparison of the charges obtained here via the Noether method with those given by the holographic energy-momentum tensor and electric current derived from the renormalized action. This would generalize the results of~\cite{Papadimitriou:2005ii} to theories including Chern-Simons terms and higher-derivative couplings.

Finally, we comment on higher-derivative supergravity in four dimensions. 
We point out that the part of the four-derivative action given in \cite{Bobev:2020egg,Bobev:2021oku} which contributes non-trivially to the equations of motion is equivalent, through a perturbative field redefinition, to the original two-derivative action with a corrected Newton's constant. This observation streamlines the derivation of the boundary terms as well as of the four-derivative corrections to the conserved charges and the black hole thermodynamics.

Another interesting direction for future work will be the extension of our results to supergravity theories with matter couplings, in particular including multiple Abelian gauge fields. This would allow to address higher-derivative corrections to black holes and other gravitational solutions carrying multiple electric charges. While extending our examination of Wald's formalism for conserved charges to Chern-Simons terms involving multiple gauge fields is straightforward, we anticipate that finding the correct generalization of the boundary terms to be added to the bulk action will be technically involved, as generically the scalar fields present in the supergravity vector multiplets run in the solution, and the boundary terms depend on the details of their asymptotic behavior. We expect that requiring AlAdS asymptotics for all fields in the solution will make the analysis more manageable, in analogy with the scalar-less setup discussed in this paper.

We also note that our analysis of conserved charges does not rely on the linearization in the higher-derivative corrections, hence it would be applicable more generally if a gravitational action valid at non-linear order in the corrections, such as the Gauss-Bonnet action and its generalizations, is given.

The plan of the paper is as follows. In Section~\ref{sec:holographicrenormalization} we discuss the boundary terms required by the variational problem and by holographic renormalization in five dimensions. In Section~\ref{sec:conservedchargesgeneral} we define the conserved charges, and in Section~\ref{sec:aladsthermodynamics} we implement these definitions into black hole thermodynamics by proving the first law and the quantum statistical relation. In Section~\ref{sec:corrections_GR} we apply our general formulae and compute the corrected charges of the supersymmetric AdS$_5$ black hole. Section~\ref{sec:4d_remarks} contains our comments on higher-derivative supergravity in four dimensions.

\section{Boundary terms in five-dimensional higher-derivative supergravity}\label{sec:holographicrenormalization}

In this section, we work in five dimensions and consider a four-derivative theory for the metric and an Abelian gauge field, including Chern-Simons terms. Our focus is on the supergravity action given in \cite{Cassani:2022lrk}, however we also make some general considerations. 
We discuss the boundary terms that are needed in order to obtain a well-defined variational problem and implement holographic renormalization in AlAdS solutions. 
 We will show that the boundary terms we used in \cite{Cassani:2022lrk} are indeed sufficient.

\subsection{Setting the stage}

In~\cite{Cassani:2022lrk} we gave a rather simple action for the bosonic sector of four-derivative minimal gauged supergravity, where the dynamical fields are the metric $g_{\mu\nu}$ and an Abelian gauge field $A_\mu$, and the four-derivative couplings are treated as a perturbation controlled by a parameter $\alpha$ with the dimension of a length squared. This action was obtained by starting from off-shell supergravity coupled to higher-derivative supersymmetric invariants \cite{Hanaki:2006pj,Ozkan:2013nwa,Butter:2014xxa,Gold:2023dfe}, integrating out the auxiliary fields and using field redefinitions to simplify the result (see also~\cite{Cremonini:2008tw,Baggio:2014hua,Bobev:2021qxx,Liu:2022sew} for related results). These manipulations were performed at {\it linear} order in $\alpha$, so the final result is supersymmetric up to higher-order terms. It reads:
\begin{equation}
\begin{aligned}\label{eq:action5D}
S_{\rm bulk}\,=&\,\frac{1}{16\pi G}\int \diff^5x \,e \left\{{c}_0 R+12{c}_1g^2-\frac{{c}_2}{4}F^2-\frac{{c}_3}{12\sqrt{3}}\,\epsilon^{\mu\nu\rho\sigma\lambda}F_{\mu\nu}F_{\rho\sigma} A_{\lambda}\right.\\[2mm]
&\left.\,+\,\lambda_1 \alpha \left[{\cal X}_{\text{GB}}-\frac{1}{2}C_{\mu\nu\rho\sigma}F^{\mu\nu}F^{\rho\sigma}+\frac{1}{8}F^4-\frac{1}{2\sqrt{3}}\,\epsilon^{\mu\nu\rho\sigma\lambda}R_{\mu\nu\alpha\beta}R_{\rho\sigma}{}^{\alpha\beta} A_{\lambda}\right]\right\}\, ,
\end{aligned}
\end{equation}
where ${\cal X}_{\text{GB}}=R_{\mu\nu\rho\sigma}R^{\mu\nu\rho\sigma}-4 R_{\mu\nu}R^{\mu\nu}+R^2$ is the Gauss-Bonnet combination,  $C_{\mu\nu\rho\sigma}=R_{\mu\nu\rho\sigma}-\frac{2}{3}\left(R_{\mu[\rho}g_{\sigma]\nu}+R_{\nu[\sigma}g_{\rho]\mu}\right)+\frac{1}{6}Rg_{\mu[\rho}g_{\sigma]\nu}$ is the Weyl tensor, $F^2= F_{\mu\nu}F^{\mu\nu}$ and $F^4= F_{\mu\nu}F^{\nu\rho}F_{\rho\sigma}F^{\sigma\mu}$. The $c$-coefficients are\footnote{These $c_k$ coefficients were denoted $\tilde{c}_k$ in~\cite{Cassani:2022lrk}.}
\be\label{eq:ccoeffs}
c_0 =1+ 4\lambda_2\alpha g^2\,,\,\,\,c_1 =1-(10\lambda_1 - 4 \lambda_2) \alpha g^2\,,\,\,\, c_2 = 1+ 4(\lambda_1 + \lambda_2) \alpha g^2\,,\,\,\, c_3 = 1 -4(3\lambda_1 - \lambda_2)\alpha g^2\,,
\ee
and $\lambda_1,\,\lambda_2$ are arbitrary dimensionless parameters. While in the original off-shell supergravity both these parameters control four-derivative invariants, after the massaging made in~\cite{Cassani:2022lrk} only $\lambda_1$ remains to control four-derivative couplings, while $\lambda_2$ appears as a correction to the gravitational coupling constant, $G$.  In fact it will be useful to introduce the effective gravitational coupling constant, 
\be\label{eq:Geff}
\frac{1}{G_{\rm eff}} \,\equiv\, \frac{c_0}{G} \,=\, \frac{1}{G}(1+4\lambda_2\alpha g^2)\,.
\ee

The bulk action should be supplemented by boundary terms. In order to discuss such terms we need to first regulate the spacetime:
%Indeed we are interested in asymptotically locally AdS (AlAdS) solutions, where the boundary really is a conformal boundary and the bulk metric diverges while approaching it.
we assume this %$\mathcal M$ 
 is foliated by hypersurfaces diffeomorphic to the conformal boundary, 
 %$\partial\mathcal M$ 
 and introduce adapted coordinates $x^\mu = \{x^i,z\}$, $i=0,1,2,3$, such that the hypersurfaces are labelled by $z$ and the conformal boundary is found at $z=0$. Adopting a Fefferman-Graham gauge, we can write the metric and gauge field in the form
\begin{equation}
\diff s^2\, =\, \ell^2\,\frac{\text dz^2}{z^2} + h_{ij}(x,z)\,\text d x^i \text dx^j\,\,, \qquad A= A_i(x,z)\, \diff x^i\,\,,
\label{eq:FGgauge}\end{equation}
where the parameter $\ell$ is identified with the radius of the AdS solution.
Then, $h_{ij}$ and $A_i$ are the induced fields on the hypersurface at fixed $z$,
and the outward-pointing normal vector with unit norm reads
\begin{equation}
n = -\frac{z}{\ell}\,\partial_z\,\,.
\label{eq:normalvector}\end{equation}
 We denote by $\mathcal{R}_{ij}$ and $\mathcal{R}$ the Ricci tensor and Ricci scalar of $h_{ij}$, respectively, while the extrinsic curvature $K_{ij}$ of the hypersurface, its trace $K$ and the normal derivative of the gauge field $E_i$ are given by
\be\label{eq:extrinsiccurv}
K_{ij} = -\frac{z}{2\ell}\,\partial_z h_{ij}\,\,,\qquad K = h^{ij} K_{ij}\,\,,\qquad
E_i = n^\mu F_{\mu i} = -\frac{z}{\ell}\partial_z A_i\,\,.
\ee

We next assume the bulk spacetime terminates at the cutoff hypersurface $z=\epsilon$, and impose Dirichlet boundary conditions for the fields. In the gauge \eqref{eq:FGgauge}, this means that the field variations satisfy
 \be\label{eq:dirichlet}
\delta h_{ij}= \delta A_i=0 \quad\text{for}\ z=\epsilon\,. 
 \ee
 
In~\cite{Cassani:2022lrk}, we argued that the boundary terms to be added to the bulk action~\eqref{eq:action5D} are:
\begin{equation}
\begin{aligned}\label{GHY_bdy_term}
S_{\text{GHY}}&=\frac{1+4\lambda_2\alpha g^2}{8\pi G}\int_{z=\epsilon}
 \diff^4x\sqrt{-h}\, {K} \\[1mm]
& -\,   \frac{\alpha\lambda_1}{4\pi G}\int_{z=\epsilon}%_{\partial {\cal M}}
 \diff^{4}x\,\sqrt{-h} \,\left[\frac{1}{3}K^3-K K_{ij}K^{ij}+\frac{2}{3}K_{ij}K^{jk}K_{k}{}^{i}+ 2 \Big({\cal R}_{ij}-\frac{1}{2}h_{ij}{\cal R} \Big)K^{ij}\right],
\end{aligned}
\end{equation}
and
\begin{equation}\label{counterterms}
S_{\text{ct}}= -\frac{1}{8\pi G}\int_{z=\epsilon}%_{\partial {\cal M}}
  \diff^4x \sqrt{-h} \, \left(3g \mu_1 +\frac{\mu_2}{4g}\,{\cal R}\right)\, ,
\end{equation}
where the coefficients $\mu_1$ and $\mu_2$ read
\begin{equation}\label{eq:mus}
\mu_1\,=\, 1+ \left(-\frac{16}{3}\lambda_1+4\lambda_{2}\right)\alpha g^2\,, \hspace{1cm}\mu_2\,=\,  1+ \left(8\lambda_1+4\lambda_{2}\right)\alpha g^2\, .
\end{equation}
Here, $S_{\rm GHY}$ represents the generalized Gibbons-Hawking-York (GHY) boundary terms needed for the variational principle with Dirichlet boundary conditions to be well-posed asymptotically,
while $S_{\text{ct}}$ denotes the local counterterms that remove the divergences from the action. Note that the counterterms take the same form as in the two-derivative theory, however their coefficients receive corrections at order $\alpha$.
After having added the boundary terms, the cutoff is removed by taking $\epsilon \to 0$. The renormalized on-shell action $S_{\rm ren}$ is then defined as
\be
S_{\rm ren} \,=\, \lim_{\epsilon\to 0} \left(S_{\rm bulk} + S_{\rm GHY} + S_{\rm ct} \right)\,.
\ee

While the derivation given in~\cite{Cassani:2022lrk} was based on the specific black hole solution studied there and was therefore only partial, below we present a more complete derivation of the boundary terms \eqref{GHY_bdy_term}, \eqref{counterterms}, valid for general AlAdS solutions.

It will be useful to recall that the equations of motion defined by the action above were spelled out in Appendix C of~\cite{Cassani:2022lrk}. Studying these equations, one finds that the AdS radius $\ell$ is corrected from its two-derivative value $\ell_{2\partial}=\frac{1}{g}$ as
\be
\ell \,=\, \frac{1}{g} \left(1+ 4\lambda_1\alpha g^2 \right)\,.
\ee
We will also need the dictionary with the dual SCFT anomaly coefficients~$\aa$,~$\cc$. Studying the holographic Weyl and R-symmetry anomalies defined by the action~\eqref{eq:action5D}, it was obtained in~\cite{Cassani:2022lrk}:
\begin{equation}
\begin{aligned}\label{dictionary_ac}
\aa \,&=\, \frac{\pi}{8Gg^3} \left( 1 +4\lambda_2 \alpha g^2  \right)\,,\\[1mm]
\cc \,&=\, \frac{\pi}{8Gg^3} \left( 1 +4(2\lambda_1+\lambda_2)\alpha g^2 \right) \,.
\end{aligned}
\end{equation}

\subsection{The variational problem}\label{sec:variational_pb}

We justify here the GHY terms given in \eqref{GHY_bdy_term}. The first line is just the usual GHY term ensuring that the two-derivative part of the bulk action---that is the first line of~\eqref{eq:action5D}---has a good Dirichlet variational problem. The second line of \eqref{GHY_bdy_term} is the boundary term introduced in~\cite{Myers:1987yn}, fixing the Dirichlet variational problem for the Gauss-Bonnet term appearing in our action. We now argue that this is all that we need as long as the variational problem is considered asymptotically in an AlAdS spacetime.

It is well-known that in higher-derivative gravity one cannot just impose Dirichlet boundary conditions, as the field equations are of order higher than two (the Gauss-Bonnet term is an exception as it generates second-order equations). 
 Identifying the additional boundary data to be specified is a non-trivial task, and different strategies have been proposed in the literature. In~\cite{Deruelle:2009zk}, and more recently~\cite{Erdmenger:2022nhz}, an approach involving auxiliary fields has been developed. The auxiliary fields are useful to identify the additional degrees of freedom that are propagated by the field equations, however this does not evade the need to fix more boundary data than the metric and (if present) the gauge field. Another approach consists of treating the higher-order terms as corrections to the two-derivative theory. Doing so one only has to study two-derivative equations, as at each order in the corrections the higher-derivative contributions are evaluated on a solution to the lower-order equations. In this perturbative approach, therefore, it should be possible to construct a generalized GHY boundary term order by order in the perturbative expansion.
 See e.g.~\cite{Smolic:2013gz} for a thorough discussion in the context of pure gravity.
 In the following we will show that when a gauge field is coupled to the metric curvature, the corresponding variational problems get combined, and this affects the GHY term. This issue was already encountered in~\cite{Cremonini:2009ih}. Instead of invoking the perturbative expansion in the higher-order corrections, we will prefer to follow a different strategy: we will assume the asymptotic behavior of the fields is the one taken in AlAdS spacetimes, and show that the terms that make the variational principle problematic are suppressed asymptotically.

Let us consider the term in the bulk action~\eqref{eq:action5D} coupling the Weyl tensor to the gauge field strength, $C_{\mu\nu\rho\sigma}F^{\mu\nu}F^{\rho\sigma}$. We actually perform a slightly more general analysis and consider generic four-derivative couplings of the Riemann curvature with $F_{\mu\nu}$,
\begin{equation}
S_{4\partial,F} \,=\,  \int\text d^5 x\, e\left(u_1 R_{\mu\nu\rho\sigma} F^{\mu\nu} F^{\rho\sigma} + u_2 R_{\mu\nu}F^{\mu\rho}F_{\rho}{}^{\nu} + u_3R F^2\right)\,\,,
\label{eq:u123combination}\end{equation} 
where $u_1, u_2, u_3$ are arbitrary coefficients. The combination that gives the $C_{\mu\nu\rho\sigma}F^{\mu\nu}F^{\rho\sigma}$ term in our bulk action is
\be\label{values_u123}
u_2 = \frac{4}{3}u_1\,,\qquad u_3 = \frac{1}{6}u_1\,,\qquad \text{with}\ u_1= - \frac{\lambda_1\alpha}{32\pi G}\,\,.
\ee 
The metric variation of \eqref{eq:u123combination} reads:
\begin{equation}
\begin{aligned}
\delta S_{4\partial,F} \,=\,
\int\text d^5 x \, e \left(u_1 \delta R^\mu_{\,\,\,\nu\rho\sigma} F_\mu^{\,\,\,\nu} F^{\rho\sigma}+u_2\,\delta R_{\mu\nu}F^{\nu\rho}F_\rho^{\,\,\,\mu} + u_3\,g^{\mu\nu}\delta R_{\mu\nu} F^2\right)+ \ldots\,,
\end{aligned}
\label{eq:variationriemann}\end{equation}
with
\begin{equation}
\delta R^\mu_{\,\,\,\nu\rho\sigma} = 2\nabla_{[\rho}\delta\Gamma^\mu_{\,\,\,\sigma]\nu}\,\,,\qquad \delta\Gamma^\mu_{\sigma\nu} = \frac{1}{2}g^{\mu\rho}\left(2\nabla_{(\sigma}\delta g_{\nu)\rho} - \nabla_\rho\delta g_{\nu\sigma} \right)\,\,.
\label{eq:riemanntometric}\end{equation}
The ellipsis comprises terms that vanish upon imposing Dirichlet boundary conditions for the metric. In the next formulae, the ellipsis will continue to denote such terms, as well as all bulk terms that are not total derivatives and lead to the bulk equations of motion.
Plugging the first of (\ref{eq:riemanntometric}) into (\ref{eq:variationriemann}) 
and integrating by parts, we are left with the following surface terms
\begin{equation}
\begin{aligned}
\delta S_{4\partial,F} \,=\, \int\text d^5 x \,e\, \nabla_\rho\Big[ 2 u_1 \left(\delta\Gamma^\mu_{\,\,\,\nu\lambda} F_\mu^{\,\,\,\nu} F^{\rho\lambda}\right)+\delta\Gamma^\rho_{\,\,\,\mu\nu}\left(u_2\,F^{\nu\lambda}F_\lambda^{\,\,\,\mu}+ u_3\,g^{\mu\nu}F^2\right) - \\
-\delta\Gamma^\nu_{\,\,\,\nu\mu} \left(u_2\,F^{\rho\lambda}F_\lambda^{\,\,\,\mu} + u_3\,g^{\mu\rho} F^2\right)\Big]+\ldots\,\,.
\end{aligned}
\label{eq:variationriemann3}\end{equation}
Passing to the boundary integral on the cutoff hypersurface and using the second of \eqref{eq:riemanntometric} we obtain
\begin{equation}
\begin{aligned}
\delta S_{4\partial,F}  = \!\int%_{\Sigma_\epsilon}
 \text d^4 x\sqrt{-h}\,n_\mu\Big[2u_1 \nabla_\nu\delta g_{\lambda\rho} F^{\rho\nu}F^{\mu\lambda} -\frac{1}{2}g^{\nu\sigma}\nabla_\rho\delta g_{\nu\sigma} \left(u_2\, F^{\mu\lambda}F_\lambda{}^{\rho}+ u_3\,g^{\mu\rho} F^2 \right)
 \\[1mm]
\ + \Big(g^{\mu\rho}\nabla_\sigma\delta g_{\nu\rho}-\frac{1}{2}\nabla^\mu\delta g_{\sigma\nu}\Big)\Big(u_2 \,F^{\nu\lambda}F_\lambda^{\,\,\,\sigma} + u_3\, g^{\sigma\nu}F^2 \Big) \Big] + \ldots \,,
\end{aligned}
\label{eq:variationriemann4}\end{equation}
where we recall that $n^\mu$ is the  unit normal vector.
We now assume the Fefferman-Graham gauge \eqref{eq:FGgauge} and recall the definitions \eqref{eq:extrinsiccurv} for the normal derivatives of the fields.
It is then straightforward to show that \eqref{eq:variationriemann4} becomes
\begin{equation}
\begin{aligned}
\delta S_{4\partial,F} \,=\, 
  \int%_{\Sigma_\epsilon} 
  \text d^4x\sqrt{-h}&\Big[
-\left(4u_1 -u_2\right)  \delta K_{ij} E^{i} E^{j} + \left(u_2-4u_3\right) h^{kl} \delta K_{kl}  E^i E_i \\[1mm]
&\,\ -u_2 \,\delta K_{ij} F^{ik} F_k^{\,\,\,j} - 2u_3\, h^{kl} \delta K_{kl} F^{ij}F_{ij}\Big]+ \ldots\,\,,
 \end{aligned}
\label{eq:GHY0}\end{equation}
where we have used the relation between variations of the induced metric and variations of the extrinsic curvature,
\begin{equation}
n^\mu \nabla_\mu \delta h_{ij} \,=\,2\delta K_{ij} \,\,.
\end{equation}
It follows that the generalized GHY term yielding a good Dirichlet variational problem for the metric with the action \eqref{eq:u123combination} is 
\begin{equation}
S_{{\rm GHY},F}= \int \diff^4x\sqrt{-h}\Big[ (4u_1 -u_2) K_{ij}E^i E^j - (u_2-4u_3) K E^{i} E_i + u_2 K_{ij}F^{ik}F_k^{\,\,\,j}  + 2u_3 K F^{ij}F_{ij}\Big].
\label{eq:GHY1}
\end{equation}
In particular, using the values \eqref{values_u123} for the coefficients, we obtain the GHY term for the $C_{\mu\nu\rho\lambda}F^{\mu\nu}F^{\rho\lambda}$ term appearing in our bulk supergravity action, 
\begin{equation}\label{eq:GHY_Weyl}
S_{\rm GHY,Weyl}= -\frac{\lambda_1\alpha}{48\pi G}\int \text d^4x\sqrt{-h}\left[4K_{ij}E^i E^j -  K E^{i} E_i  + 2K_{ij}F^{ik}F_k{}^{j} +\frac{1}{2} K F^{ij}F_{ij}\right]\,\,.\end{equation}

While this fixes the Dirichlet variational principle for the metric, it spoils the 
one for the gauge field, which was so far well-posed. Indeed, \eqref{eq:GHY1} and  \eqref{eq:GHY_Weyl} involve the normal derivatives $E_i$ of the gauge field, that are not fixed by Dirichlet boundary conditions.
 As anticipated, we circumvent this issue by assuming that the fields respect the conditions for an AlAdS solution.
 In the coordinates specified by \eqref{eq:FGgauge}, this means that the induced fields admit the following small-$z$ expansion:
\begin{equation}
\begin{aligned}
h_{ij}(x,z) &= \frac{1}{z^2}\left[h_{ij}^{(0)}(x) + z^2h_{ij}^{(2)}(x) + z^4\left( h_{ij}^{(4)}(x) + \tilde h_{ij}^{(4)}(x)\,\log z\right) + \mathcal O(z^5)\right] \,\,,\\[1mm]
A_i(x,z) &= A_i^{(0)}(x) + z^2 \left(A_i^{(2)}(x) + \tilde A_i^{(2)}(x)\,\log z\right) + \mathcal O(z^3)\,\,. 
\end{aligned}
\label{eq:expansion}\end{equation}
This implies the following leading behaviour of the intrinsic curvatures of  boundary tensors,
\be
F_{ij} = \mathcal O(1)\,\,,\ \quad\nabla_i= \mathcal O(1)\,\,,\ \quad
\mathcal R^i_{\,\,\,jkl} =\mathcal O(1)\,\,,\ \quad \mathcal R_{ij} =\mathcal O(1)\,\,,\ \quad \mathcal R = \mathcal O(z^2)\,\,,
\label{eq:intrinsictensors}
\ee
while the extrinsic curvatures behave as
\begin{equation}
\begin{aligned}
& K_{ij} = \frac{1}{z^2}h_{ij}^{(0)} + \mathcal O(1)\,\,,\qquad
\nabla_i K_{jk} = \mathcal O(1)\,\,,\qquad K = 4+ \mathcal{O}(z^2)\,\,,\\[1mm]
&E_i =  -\frac{z^2}{\ell} \left(2A_i^{(2)}+ \tilde A_i^{(2)}\,(1+2\log z) \right) +\mathcal{O}(z^3) \,\,.
\label{eq:extrinsictensors}
\end{aligned}
\end{equation}
Using these expansions, it is easy to see that the terms in \eqref{eq:GHY1} (and hence \eqref{eq:GHY_Weyl}) involving $E_i$ are suppressed in the limit $\epsilon \to 0$ that removes the radial regulator. Therefore, even though the Dirichlet problem for the gauge field is ill-defined for finite values of the cutoff, the issue disappears in the limit where the latter is removed. We conclude that the Dirichlet problem for the action $S_{4\partial,F}+ S_{{\rm GHY},F}$ given in  \eqref{eq:u123combination}, \eqref{eq:GHY1} is well-posed {\it asymptotically} in AlAdS spacetimes.

We further observe that while the terms involving $E_i$ are suppressed, the last two terms in \eqref{eq:GHY1} are finite in the asymptotic expansion, hence generically they contribute to the renormalized action. However, for the specific case of the $C_{\mu\nu\rho\sigma}F^{\mu\nu}F^{\rho\sigma}$ coupling appearing in our bulk supergravity action, the whole GHY term is suppressed asymptotically since the finite contributions from the last two terms in \eqref{eq:GHY_Weyl} add up to zero.
For this reason, we omitted the GHY term associated with $C_{\mu\nu\rho\sigma}F^{\mu\nu}F^{\rho\sigma}$ in our boundary terms \eqref{GHY_bdy_term}.

An analogous argument was previously used in~\cite{Landsteiner:2011iq} 
to treat the gauge-gravitational Chern-Simons term, $\epsilon^{\mu\nu\rho\sigma\lambda}R_{\mu\nu\alpha\beta}R_{\rho\sigma}{}^{\alpha\beta} A_{\lambda}$, which also appears in our bulk action~\eqref{eq:action5D}. By requiring that the action reproduces the gauge-gravitational anomaly on a general hypersurface a boundary term was derived in~\cite{Landsteiner:2011iq}, however it contains normal derivatives of the extrinsic curvature and does not fix the Dirichlet variational problem for the metric.\footnote{Nevertheless, it simplifies the expression for the action when this is expressed in ADM variables by cancelling a bulk total derivative. In Sec.~\ref{sec:HJmethod} we show that this is also the case for our boundary term~\eqref{eq:GHY1}.} This issue is avoided by noting that the terms not fixed by the Dirichlet boundary conditions are suppressed asymptotically for field configurations respecting the expansion~\eqref{eq:expansion}. 
While we refer to~\cite{Landsteiner:2011iq} for details,\footnote{See also~\cite{Grumiller:2008ie} for a study in four dimensions, where the gauge field is replaced by an axion-like scalar.} here we simply conclude that it is not necessary to associate a boundary term to the gauge-gravitational Chern-Simons term for our purposes.

Noting that the Dirichlet problem for $F^4$ is well-defined with no boundary term added, the analysis of the variational problem for the bulk action~\eqref{eq:action5D} is then complete. 

This concludes our argument for the GHY boundary term given in  \eqref{GHY_bdy_term}. In Section~\ref{sec:HJmethod} we will derive the boundary counterterms \eqref{counterterms} that implement holographic renormalization by removing the divergences from the on-shell action. 

\subsubsection{A fully well-defined Dirichlet problem}
As an aside to our main discussion, we observe that there exists a special way to couple the Riemann curvature with the gauge field strength $F_{\mu\nu}$ such that the Dirichlet problem is well-defined both for the metric and the gauge field at finite values of the radial cutoff $\epsilon$. This corresponds to choosing the coefficients in \eqref{eq:u123combination} as
\be
u_2 = 4u_1\,,\qquad  u_3 = u_1\,,
\ee
so that the GHY term \eqref{eq:GHY1} does not contain normal derivatives of the gauge field, $E_i$.
The corresponding combination of bulk terms is
\begin{equation}
\begin{aligned}
S_{4\partial,F} \,&=\, u_1\int \diff^5x \,e \left(R_{\mu\nu\rho\sigma} F^{\mu\nu} F^{\rho\sigma} + 4 R_{\mu\nu}F^{\mu\rho}F_{\rho}{}^\nu + R F^2\right)\\
\,&=\, -  u_1\int \diff^5x \,e \,\epsilon^{\lambda\mu\nu\rho\sigma} R_{\mu\nu}{}^{\mu'\nu'}\epsilon_{\lambda\mu'\nu'\rho'\sigma'} F_{\rho\sigma}F^{\rho'\sigma'}\,,
\label{eq:specialcombination}
\end{aligned}
\end{equation} 
and the GHY term reads
\begin{equation}
S_{{\rm GHY},F} \,=\, 2u_1\int\text d^4x\sqrt{-h}\left( 2K_{ij}F^{ik}F_k^{\,\,\,j} +KF_{ij}F^{ij} \right)\,\,.
\label{eq:genGHY_F}
\end{equation}
In this case, the variational problems for the metric and the gauge field are both well-posed. Indeed, variations of $S_{4\partial,F} + S_{{\rm GHY},F}$ with respect to the gauge field take the schematic form
\begin{equation}
\int\text d^5 x\, e\left(\text{vector EoM} \right)^i\delta A_i + \int\text d^4x\sqrt{-h} \left(\text{bdy terms}\right)^i\delta A_i 
+\int\text d^4x\sqrt{-h}\, \frac{\delta S_{\rm GHY}}{\delta F_{ij}}2\nabla_i\delta A_j\,\,.
\label{eq:variationgaugefield}\end{equation}
The second piece denotes boundary terms emerging from the bulk action when integrating by parts so as to obtain the vector equations of motion (EoM). Since the bulk action does not involve terms with derivatives of the field strength, these boundary terms are simply proportional to $\delta A_i$. Moreover, since the GHY term (\ref{eq:genGHY_F}) does not contain normal components of the field strength, the term in (\ref{eq:variationgaugefield}) involving its variation vanishes identically when we impose the Dirichlet boundary conditions \eqref{eq:dirichlet}.

One can also check that both the metric and gauge field equations generated by this action are  of second-order, consistently with the fact that we have found a GHY term. 

We stress that the present proof does not require the fields to satisfy the AlAdS expansion~\eqref{eq:expansion}, nor it needs $\alpha$ to be small. Moreover, although we have given it for five bulk dimensions, it works in the same way in arbitrary dimension. So the sum of \eqref{eq:specialcombination} and \eqref{eq:genGHY_F} provides an analog of the Gauss-Bonnet term and its corresponding GHY term for the couplings of the Riemann curvature with a gauge field.

\subsection{Holographic renormalization}\label{sec:HJmethod}

\subsubsection{Brief review of the Hamilton-Jacobi method}\label{sec:hamiltonjacobigeneral}

After discussing the variational principle, we can turn to the cancellation of the divergences in the on-shell action by means of holographic renormalization. We will adopt the approach based on the Hamilton-Jacobi method~\cite{deBoer:1999tgo,Fukuma:2000bz,Martelli:2002sp} -- see also~\cite{Papadimitriou:2004ap} for a related method, and~\cite{Elvang:2016tzz,Papadimitriou:2016yit} for nice recent discussions. This systematizes the analysis of counterterms and appears particularly convenient in the presence of higher-derivative terms. Compared to existing analyses using the Hamilton-Jacobi method in higher-derivative gravity, such as~\cite{Fukuma:2001uf,Liu:2008zf}, here we include couplings to the gauge field $A$.

The idea is to solve perturbatively the Hamilton-Jacobi equation
\begin{equation}\label{eq:hamiltonjacobi}
  {\cal H}\left[\frac{16\pi G_{\rm eff}}{\sqrt{-h}}\frac{\delta S_{\text{reg}}}{\delta h_{ij}}\,,\ \frac{16\pi G_{\rm eff}}{\sqrt{-h}}\frac{\delta S_{\text{reg}}}{\delta A_{i}}\ ;\ h_{ij}\,,\ A_i\right]=0\,\,,
\end{equation}
where the regulated action 
\be
S_{\text{reg}}=S_{\rm bulk}+S_{\rm GHY}
\ee
 is the action evaluated on a solution in the spacetime ending at $z=\epsilon$. Here, $\mathcal{H}$ is the Hamiltonian density of the theory, where the role of time is being played by the radial coordinate $z$. Eq.~\eqref{eq:hamiltonjacobi} arises as the Hamiltonian constraint associated with reparametrization invariance. 
In terms of $S_{\text{reg}}$, the renormalized on-shell action $S_{\rm ren}$ is given by
\begin{equation}
S_{\text{ren}} \,=\, S_{\text{reg}} + S_{\rm ct}\,\,,
\label{def:onshellrenormalizedaction}\end{equation}
where it is understood that in the right hand side we neglect all terms that are suppressed in the limit $\epsilon\to 0$.
By construction this is finite, or at most logarithmically divergent in the case where a holographic Weyl anomaly is present. 
 Then Eq.~\eqref{eq:hamiltonjacobi} should be regarded as a first-order partial differential equation for $S_{\text{ren}}$ and $S_{\rm ct}$. Here we are only interested in solving the leading orders of \eqref{eq:hamiltonjacobi} near the boundary, so as to determine the local counterterms that remove the power-law divergences from the on-shell action, as well as the holographic Weyl anomaly.
 
 In order to do that we first need to identify the divergent sector of the on-shell action. It is convenient to rewrite the Lagrangian density $\mathcal{L}_{\rm reg}$, that for a $(d+1)$-dimensional spacetime is related to the regulated action as $S_{\rm reg} = \frac{1}{16\pi G_{\rm eff}}\int\text d^{d+1}x\,e\,\mathcal L_{\rm reg}$, in terms of the fields $h_{ij}$ and $A_i$ induced on the hypersurface at fixed $z$, making use of Gauss-Codazzi relations. After doing that, the Lagrangian density can be expanded as
\begin{equation}
\mathcal L_{\rm reg} = \mathcal L^{(0)} + \mathcal L^{(2)} + ... + \mathcal L^{(d)} + ...\,\,,
\end{equation}
where every term $\mathcal L^{(2k)}$ comprises terms whose leading-order behavior near the boundary is $\mathcal{O}(z^{2k})$ when the asymptotic expansion of the fields is substituted
%[asymptotic behavior = 2( n. of $h^{-1}$ + n. of $E_i$ - n. of $K_{ij}$)]. 
 (in the case of pure gravity this expansion is equivalent to a derivative expansion).
The leading terms in the expansion, starting from $\mathcal L^{(0)}$ up to $\mathcal L^{(d-2)}$, when integrated, yield power-law divergences in $S_{\rm reg}$. These divergences should be cancelled by the counterterms $S_{\rm ct}$, as requested by \eqref{def:onshellrenormalizedaction}. For even $d$ there can also be a term $\mathcal L^{(d)}$ in the Lagrangian, which gives a logarithmically divergent contribution to the action. The higher-order terms in the expansion of $\mathcal L_{\rm reg}$ are sufficiently suppressed asymptotically and do not affect the divergent sector, hence can be neglected for the purposes of the present discussion. 

Implementing the Hamilton-Jacobi method asymptotically then consists of the following steps.

\paragraph{Introducing the Hamiltonian.} First, we need to determine the Hamiltonian density $\mathcal H$, after rewriting $\mathcal L_{\rm reg}$ using the Gauss-Codazzi relations. 
 This is defined as
\begin{equation}\label{def:hamiltonian}
\mathcal H \,=\,   \frac{16\pi G_{\rm eff}}{\sqrt{-g}}\left(\frac{\delta S_{\rm reg}}{\delta \dot h_{ij}}\,\dot h_{ij} + \frac{\delta S_{\rm reg}}{\delta \dot A_i}\,\dot A_i\right) -\mathcal L_{\rm reg}\,\,,
\end{equation}
where the dot means a derivative with respect to $z$.
The canonical momenta are given by
\begin{equation}
\pi^{ij} \,\equiv\,  -\frac{16\pi G_{\rm eff}}{\sqrt{-h}}\,\frac{\delta S_{\rm reg}}{\delta \dot h_{ij}} =\frac{16\pi G_{\rm eff}}{2\sqrt{-g}}\frac{\delta S_{\rm reg}}{\delta K_{ij}}
\,\,,\qquad  \
\pi^i \,\equiv\,  -\frac{16\pi G_{\rm eff}}{\sqrt{-h}}\frac{\delta S_{\rm reg}}{\delta \dot A_i} =\frac{16\pi G_{\rm eff}}{\sqrt{-g}}\frac{\delta S_{\rm reg}}{\delta E_i}
\,\,,
\label{def:conjugatemomenta}\end{equation}
hence, recalling \eqref{eq:extrinsiccurv}, the Hamiltonian density becomes 
\begin{equation}
\mathcal H \,=\, 2\pi^{ij} K_{ij} + \pi^i E_i-\mathcal L_{\rm reg}\,\,.
\label{eq:hamiltonian}\end{equation}
The momenta may be expanded as
\begin{equation}
\pi^{ij} = \pi^{ij}_{(2)} + \pi^{ij}_{(4)} + ... + \pi^{ij}_{(2+d)} + ... \,\,,
\end{equation}
and similarly for $\pi^i$. Here, each term is associated to a specific term in the expansion of $\mathcal L_{\rm reg}$. In particular, for even $d$ there is a term $\pi^{ij}_{(2+d)}$ coming from variations of $\mathcal L^{(d)}$.
It follows that the Hamiltonian density can be expanded as
\begin{equation}
\mathcal H \,=\, \mathcal H^{(0)} + \mathcal H^{(2)} + ... + \mathcal H^{(d)} + ... \,\,.
\end{equation}
%where the ellipsis indicates terms coming from the terms in the expansion of $\mathcal L_{\rm reg}$ that are sufficiently suppressed asymptotically so as not to lead to divergences in the action.

\paragraph{Hamiltonian in terms of momenta.} The definition \eqref{def:conjugatemomenta} for the momenta can be inverted in order to express the extrinsic tensors $K_{ij}$ and $E_i$ as functions of the momenta $\pi^{ij}$ and $\pi^i$. In this way we obtain the Hamiltonian \eqref{eq:hamiltonian} as a function of the momenta, $\mathcal H = \mathcal H(\pi^{ij}, \pi^i; h_{ij} , A_i)$.

We can discuss the dependence of the different terms in the asymptotic expansion of the Hamiltonian on the different terms in the momenta. 
We notice that $\mathcal H^{(0)}$ can only depend on the momentum conjugate to the metric via the trace of $\pi^{ij}_{(2)}$, since it is the only possible combination with the correct order in the asymptotic expansion. Similarly, $\mathcal H^{(2k)}$ will depend on the trace of $\pi^{ij}_{(2+2k)}$ and some combinations of the momenta $\pi^{ij}_{(n)}$ with $n<2k+2$. 
Finally, $\mathcal H^{(d)}$ depends on the trace of $\pi^{ij}_{(2+d)}$ and on the lower-order terms.

\paragraph{Counterterms and Weyl anomaly.} The Hamilton-Jacobi method now prescribes to replace the momenta appearing in the Hamiltonian \eqref{eq:hamiltonian} with the functional derivatives,
\begin{equation}
\pi^{ij} = \frac{16\pi G_{\rm eff}}{\sqrt{-h}}\left(-\frac{\delta S_{\rm ren}}{\delta h_{ij}} + \frac{\delta S_{\rm ct}}{\delta h_{ij}}\right)\,,\qquad\quad
\pi^{i} = \frac{16\pi G_{\rm eff}}{\sqrt{-h}}\left(-\frac{\delta S_{\rm ren}}{\delta A_i} + \frac{\delta S_{\rm ct}}{\delta A_i}\right)\,.
\label{eq:momentahj}\end{equation}
This yields the Hamilton-Jacobi equation \eqref{eq:hamiltonjacobi}, that can be solved order by order in the expansion of $\mathcal{H}$. 
The momenta can be split in two contributions
\begin{equation}
\pi = \pi^{(\rm ren)} + \pi^{\rm (ct)} \,\,,
\end{equation}
where $\pi^{(\rm ren)}$ contains the terms contributing to the renormalized action, while $\pi^{\rm (ct)}$ contributes to power-law divergences.
 We can now introduce an ansatz for the counterterm action $S_{\rm ct}$ as a linear combination of all local covariant boundary terms that diverge asymptotically, with coefficients to be determined. Expanding the Hamilton-Jacobi equation \eqref{eq:hamiltonjacobi} order by order gives a set of equations $\mathcal{H}^{(n)}=0$. Those constraining the power-law divergent sector of $S_{\rm reg}$ are the leading-order ones up to $\mathcal H^{(n)}$ with $n<d$. These just depend on $\pi^{\rm (ct)}$ and allow us to extract the coefficients fixing $S_{\rm ct}$. When $d$ is even we should also consider  $\mathcal H^{(d)}=0$: from this equation we can extract the holographic Weyl anomaly, since it depends on $h_{ij}\frac{1}{\sqrt{-h}}\frac{\delta S_{\rm ren}}{\delta h_{ij}}$ through the trace of $\pi^{ij}_{(2+d)}$ \cite{Martelli:2002sp,Fukuma:2001uf}.

\subsubsection{Implementing the procedure for our action}
\label{sec:hamiltonjacobi5D}

We now apply the procedure reviewed above to five-dimensional minimal gauged supergravity with four-derivative couplings.

\paragraph{Divergent sector.} In order to obtain the Hamiltonian $\mathcal H$ as in \eqref{eq:hamiltonian}, we first express the Lagrangian density $\mathcal L_{\rm reg} = \mathcal L_{\rm bulk} + \mathcal L_{\rm GHY}$ using the Gauss-Codazzi relations. 
In the gauge specified by \eqref{eq:FGgauge}, the non-vanishing components of the Christoffel symbol read
\begin{equation}
\begin{aligned}
&\qquad\quad\Gamma^i_{\,\,\,jk} = \gamma^i_{\,\,\,jk}\,\,,\qquad \Gamma^z_{\,\,\,ij} = -\frac{1}{2}g^{zz}\partial_z h_{ij}=\frac{z}{\ell} K_{ij}\,\,,\\[1mm]
&\Gamma^z_{\,\,\,zz}=\frac{1}{2}g^{zz}\partial_z g_{zz} = -z^{-1}\,\,,\quad \,\,\, \Gamma^i_{\,\,\,jz} = \frac{1}{2}h^{ik}\partial_z h_{jk}=-\frac{\ell}{z}K^i_{\,\,\,j} \,\,,
\end{aligned}
\label{eq:gamma}\end{equation}
where $\gamma$ denotes the Christoffel symbol defined from $h_{ij}$.
The non-vanishing components of the Riemann tensor are
\begin{equation}
\begin{aligned}
&R^i_{\,\,\,jkl} = \mathcal R^i_{\,\,\,jkl}- K^i_{\,\,\,k}K_{jl}+ K^i_{\,\,\,l}K_{jk}\,\,,\\[1mm] 
&R^z_{\,\,\,ijk} = \frac{z}{\ell}\left(\nabla_j K_{ik} - \nabla_k K_{ij}\right)\,\,, \qquad R^z_{\,\,\,izj} = \frac{z}{\ell}\partial_zK_{ij} + K_{ik} K^k_{\,\,\,j}\,\,,
\end{aligned}
\label{eq:riemann1}\end{equation}
where $\mathcal R_{ijkl}$ and $\nabla_i$ are the Riemann tensor and the Levi-Civita connection built out of $h_{ij}$.
 %Notice that latin indices have to be raised carefully; for instance, $R^{zi}_{\,\,\,\,\,zj} = \frac{z}{\ell}\,\partial_z K^i_{\,\,\,j} - K^{ik}K_{kj}\,\,.$
The components of the Ricci tensor can be written as
\begin{equation}
\begin{aligned}
R_{ij} &= \mathcal R_{ij} - K K_{ij} +   \frac{z}{\ell}\partial_z K_{ij} + 2K_{ik} K^k_{\,\,\,j}\,\,,\\[1mm]
R_{zi} &= %R^j_{\,\,\,zji}= -h^{jk}g_{zz} R^z_{\,\,\,kji}=
  -\frac{\ell}{z}\left( \nabla_j K^j_{\,\,\,i} - \nabla_i K\right)\,\,,\\[1mm]
R_{zz} %= R^i_{\,\,\,ziz}= h^{ik} g_{zz} R^z_{\,\,\,kzi} 
 &= \frac{\ell^2}{z^2}\left(\frac{z}{\ell}\partial_z K - K_{ij}K^{ij}\right)\,\,,
\end{aligned}
\label{eq:ricci1}\end{equation}
while the Ricci scalar reads
\begin{equation}
\begin{aligned}
R &= \mathcal R + 2\frac{z}{\ell}\partial_z K -K_{ij}K^{ij}-K^2\\[1mm]
&= \mathcal R - K_{ij}^2 + K^2 + 2e^{-1}\,\partial_z\left(\sqrt{-h}\,K\right)\,\,.
\end{aligned}
\label{eq:ricciscalar}\end{equation}

We next use the Gauss-Codazzi relations above to identify the divergent part of the action $S_{\text{reg}}=S_{\rm bulk}+S_{\rm GHY}$, where $S_{\rm bulk}$ and $S_{\rm GHY}$ were given in \eqref{eq:action5D} and \eqref{GHY_bdy_term}, respectively.
For the sector made of the two-derivative action, the Gauss-Bonnet term, and the respective GHY terms, we can follow the analysis of~\cite{Liu:2008zf} -- with the  difference that we also include the gauge kinetic term $F_{\mu\nu}F^{\mu\nu}$. Using the Gauss-Codazzi relations, one finds that the whole generalized GHY term is 
cancelled by an opposite contribution coming from the bulk action, leaving us with
\begin{equation}
\begin{aligned}
 S_{\rm reg} & \supset\ \frac{1}{16\pi G_{\rm eff}} \int\text d^5 x \, e\left(\overline R + 12\hat c_1g^2 - \frac{1}{4}\hat c_2 F_{ij}^2 -\frac{1}{2}\hat c_2E_i^2\right) \\[1mm]
& +\alpha\lambda_1\Big\{\left(\overline R^2_{ijkl} - 4\overline R^2_{ij} + \overline R^2 \right)  + 4\Big[\mathcal R K^2 - \mathcal R K_{ij}^2 -4 \mathcal R_{ij} K^{ij} K +4 \mathcal R_{ij} K^{ik}K_k^{\,\,\,j} \\[1mm]
& +2 \mathcal R_{ijkl} K^{ik} K^{jl}  + 2 K^2 K_{ij}^2 -  (K_{ij}^2)^2 +2 K_{ij}K^{jk}K_{kl}K^{li}-\frac{1}{3}K^4 -\frac{8}{3}K K^i_{\,\,\,j}K^{jk}K_{ki}\Big]
\Big\},
\end{aligned}
\label{eq:gb4}\end{equation}
where the coefficients 
\begin{equation}
\hat c_1 = 1-10\lambda_1\alpha g^2\,\,, \quad\qquad \hat c_2 = 1+ 4\lambda_1\alpha g^2
\end{equation}
correspond with the coefficients $c_1$, $c_2$ given in~\eqref{eq:ccoeffs} with $\lambda_2 =0$, since the effect of this parameter has been absorbed in the redefinition \eqref{eq:Geff} of the gravitational coupling constant, $G \to G_{\rm eff}$. Furthermore, we introduced
\begin{equation}
\begin{aligned}
\overline{R}_{ijkl} &= \mathcal R_{ijkl} - K_{ik}K_{jl} + K_{il} K_{jk}\,\,,\\[1mm]
\overline{R}_{ij} &= \mathcal R_{ij} - K K_{ij} + K_{ik}K^k{}_j\,\,,\\[1mm]
\overline{R} &= \mathcal R - K^2 + K_{ij}K^{ij}\,\,.
\end{aligned}
\label{eq:overliner}\end{equation}

We now argue that the remaining part of $S_{\rm reg}$ does not lead to divergences. The only non-obvious terms are $C_{\mu\nu\rho\sigma}F^{\mu\nu} F^{\rho\sigma}$ and the gauge-gravitational Chern-Simons term in  \eqref{eq:action5D}. The latter was discussed in~\cite{Landsteiner:2011iq}: as we already noticed in Section~\ref{sec:variational_pb}, a boundary term was introduced there, but it is suppressed asymptotically in an AlAdS background and can therefore be neglected for our purposes. 
The reasoning for $C_{\mu\nu\rho\sigma}F^{\mu\nu} F^{\rho\sigma}$ is similar, as we already discussed in Section~\ref{sec:variational_pb}. Let us see this from the perspective of the Hamilton-Jacobi method. In order to do so, we  come back to the action \eqref{eq:u123combination} with generic coefficients $u_1,u_2,u_3$, for which we identified the GHY term \eqref{eq:GHY1}. Using (\ref{eq:riemann1})--(\ref{eq:ricciscalar}) and (\ref{eq:intrinsictensors}), (\ref{eq:extrinsictensors}), one can show that
\begin{equation}
\begin{aligned}
&u_1\,R_{\mu\nu\rho\lambda} F^{\mu\nu} F^{\rho\lambda} + u_2\,R_{\mu\nu}F^{\nu\rho} F_\rho^{\,\,\,\mu} + u_3\, RF^2 =\\[1mm]
& -2\left(u_1- u_2\right)K_{ik} K_{jl} F^{ij} F^{kl}- 2u_2\,K_{ij}K^{jk}F_{kl}F^{li} +u_3\left(8 K K_{ij}F^{ik}F_k^{\,\,\,j} - K_{ij}^2 F_{kl}^2 + K^2 F_{ij}^2 \right)\\[1mm]
&+e^{-1}\,\partial_z\left(\sqrt{-h}\left(u_2 K_{ij} F^{jk}F_{k}^{\,\,\,i} + 2u_3\,K F_{ij}^2 \right)\right) + \mathcal O(z^5)\,\,. 
\end{aligned}
\label{eq:weyltermgeneral}\end{equation}
The total derivative is exactly cancelled by the boundary terms in \eqref{eq:GHY1} that are not suppressed asymptotically. 
 Plugging the asymptotic behavior of the extrinsic curvature given in  (\ref{eq:extrinsictensors}), we obtain
\begin{equation}
\begin{aligned}
&u_1\,R_{\mu\nu\rho\lambda} F^{\mu\nu} F^{\rho\lambda} + u_2\,R_{\mu\nu}F^{\nu\rho} F_\rho^{\,\,\,\mu} + u_3\, RF^2 =\\
&\left(-2u_1 + 4u_2 -20 u_3\right) F_{ij}F^{ij} + e^{-1}\,\partial_z\left(\sqrt{-h}\left(-u_2 + 8u_3\right)F_{ij}F^{ij}\right) + \mathcal O(z^5)\,\,.
\end{aligned}
\label{eq:weyltermgeneral2}\end{equation}
These terms do not lead to power-law divergences in the on-shell action, but a priori the first one  yields a $\log \epsilon$ divergence when integrated in the bulk and thus contributes to the holographic Weyl anomaly, while the boundary term gives a finite contribution.

However, when the Weyl combination $C_{\mu\nu\rho\sigma}F^{\mu\nu} F^{\rho\sigma}$ is considered, corresponding to the choice $u_2 = \frac{4}{3}u_1$ and $u_3 = \frac{1}{6}u_1$,
Eq.~\eqref{eq:weyltermgeneral2} simplifies and the whole expression is further suppressed asymptotically. 

We conclude that the divergences of the regularized on-shell action are captured entirely by expression \eqref{eq:gb4}, as anticipated.\footnote{The $E_iE^i$ term appearing in \eqref{eq:gb4} does not lead to divergences either.}
 The Hamilton-Jacobi equations we have to solve, therefore, have the same form as the Einstein-Gauss-Bonnet problem discussed in~\cite{Liu:2008zf}, plus a contribution from the Maxwell kinetic term that only affects the logarithmic divergence.

\paragraph{The Hamiltonian.} Computing the momenta \eqref{def:conjugatemomenta} and evaluating the Hamiltonian density (\ref{eq:hamiltonian}), we arrive at
\begin{equation}
\begin{aligned}
\mathcal H = -\Big( \overline R + 12\hat c_1g^2 - \frac{1}{4}\hat c_2 F_{ij}^2 \Big) -\alpha\lambda_1\left( \overline R_{ijkl}^2 -4\overline R_{ij}^2 + \overline R^2\right) + ... \,\,.
\end{aligned}
\label{eq:hamiltoniandensity}\end{equation}
Here, the ellipsis comprise all the subleading contributions to the Hamiltonian, that correspond to finite contributions to the regulated action $S_{\rm reg}$, that requires no renormalization, and just affect the definition of the finite part of the renormalized action $S_{\rm ren}$. Recall that we just want to solve the Hamilton-Jacobi constraint \eqref{eq:hamiltonjacobi} asymptotically, in order to extract the counterterms action $S_{\rm ct}$ and the Weyl anomaly. 

By inverting the relation between the momenta, defined according to \eqref{def:conjugatemomenta}, and the extrinsic curvature $K_{ij}$, we obtain the following expression for the Hamiltonian~\cite{Liu:2008zf},
\begin{equation}
\begin{aligned}
\mathcal H(\pi^{ij}; h_{ij}, A_i) =&  -\Big[ \mathcal R + 12\hat c_1 g^2 -\frac{1}{2}\hat c_2 F_{ij}^2 + \pi_{ij}^2 - \frac{1}{3}\pi^2 + \alpha\lambda_1\Big( \mathcal R^2 - 4 \mathcal R_{ij}^2 + \mathcal R_{ijkl}^2 \\[1mm]
&-\frac{16}{3}\mathcal R_{ij}\pi^{ij} \pi + \frac{10}{9}\mathcal R \pi^2 - 2\mathcal R \pi_{ij}^2 + 8\mathcal R_{ij}\pi^{jk}\pi_k^{\,\,\,i} + 4\mathcal R_{ijkl} \pi^{ik}\pi^{jl} \\[1mm]
&+ 2\pi^i_{\,\,\,j}\pi^j_{\,\,\,k}\pi^k_{\,\,\,l}\pi^l_{\,\,\,i} - (\pi_{ij}^2)^2 -\frac{16}{9}\pi\pi^i_{\,\,\,j}\pi^j_{\,\,\,k}\pi^k_{\,\,\,i} + \frac{10}{9}\pi^2 \pi_{ij}^2 - \frac{11}{81}\pi^4
\Big) \Big]+\text{...}\,\,,
\end{aligned}
\label{eq:hamiltonianliuesabra}\end{equation}
where $\pi\equiv \pi^{ij}h_{ij}$, and we neglected $\mathcal O(\alpha^2)$ terms.

\paragraph{The counterterms.} The ansatz for the possible counterterms is   
\begin{equation}
S_{\rm ct} = -\frac{1}{16\pi G_{\rm eff}}\int\text d^4x\sqrt{-h}\left( W + C \mathcal R\right)\,\,,
\label{eq:countertermsactionansatz}\end{equation}
 where $W \propto g$ and  $C\propto g^{-1}$ by dimensional analysis.
 All other local, covariant terms built out of intrinsic boundary fields are either constant at the boundary, hence depend on the choice of renormalization scheme, or suppressed. 

As explained in Sec.~\ref{sec:hamiltonjacobigeneral}, the power-law divergent sector of the Hamiltonian 
 just depends on 
\begin{equation}
\frac{16\pi G_{\rm eff}}{\sqrt{-h}}\,\frac{\delta S_{\rm ct}}{\delta h_{ij}} \,=\, -\frac{W}{2}\,h^{ij} + C\left( \mathcal R^{ij} - \frac{1}{2}\mathcal R h^{ij}\right)\,\,.
\label{eq:momentahjct}\end{equation}
After substituting \eqref{eq:momentahjct} into \eqref{eq:hamiltonianliuesabra}, the first two orders of the Hamilton-Jacobi equations give algebraic conditions for determining the coefficients $W$ and $C$ introduced in \eqref{eq:countertermsactionansatz}~\cite{Liu:2008zf},
\begin{equation}
\begin{aligned}
\mathcal H^{(0)}= 12\hat c_1 g^2- \frac{1}{3}W^2  - \frac{\alpha\lambda_1}{162}W^4=0\,\,,\\
\mathcal H^{(2)}=\mathcal R\left(1- \frac{1}{3}WC + \alpha\lambda_1 \frac{1}{9}W^2\left( 1- \frac{1}{9}WC\right) \right)=0\,\,.
\end{aligned}
\label{eq:hamiltonjacobi5D}\end{equation}
These two equations are solved by 
\begin{equation}
W= 6g\left(1-\frac{16}{3}\lambda_1\alpha g^2\right)\,, \hspace{1cm} C=\frac{1}{2g}\left(1+8\lambda_1\alpha g^2\right)\,.
\label{eq:countertermscoefficients}
\end{equation}
As anticipated, these coefficients are in agreement with the boundary terms we gave in  \eqref{counterterms} \eqref{eq:mus}. 
 The theory is indeed renormalized by the same counterterms as the two-derivative action, just the respective coefficients get corrected.

\paragraph{The holographic Weyl anomaly.} The only remaining divergence in the on-shell action is the logarithmic one, which leads to the holographic Weyl anomaly. In the following we check that the Hamilton-Jacobi equation  \eqref{eq:hamiltonjacobi} gives the result expected from \cite{Fukuma:2001uf},\footnote{See also~\cite{Nojiri:1999mh,Blau:1999vz}, where the holographic Weyl anomaly for higher-derivative gravity was calculated using the original method of~\cite{Henningson:1998gx}.} with an additional contribution from the gauge field strength $F_{ij}$, which was not considered in that reference.

As we argued in Sec.~\ref{sec:hamiltonjacobigeneral}, we should consider the equation $\mathcal H^{(4)}=0$ as it depends on $h_{ij}\frac{\delta S_{\rm ren}}{\delta h_{ij}}$, that gives the trace of the holographic energy-momentum tensor. Substituting the definition \eqref{eq:momentahj} for the momenta as functional derivatives into \eqref{eq:hamiltonianliuesabra}, where now the values of $C$ and $W$ are given by \eqref{eq:countertermscoefficients}, we find that $\mathcal H^{(4)}=0$ yields
\begin{equation}
\begin{aligned}
32\pi G_{\rm eff}\left(1-4\alpha g^2\lambda_1\right) g\,{\frac{h_{ij}}{\sqrt{-h}}\frac{\delta S_{\rm ren}}{\delta h_{ij}}}=&-\frac{1}{4g^2}\left({\cal R}_{ij}{\cal R}^{ij}-\frac{1}{3}{\cal R}^2-g^2 F_{ij}F^{ij}\right)\\[1mm]
&-\lambda_1\alpha\left(C_{ijkl}C^{ijkl}-{\cal R}_{ij}{\cal R}^{ij}+\frac{1}{3}{\cal R}^2-g^2 F_{ij}F^{ij}\right)\,,
\end{aligned}
\end{equation}
where $C_{ijkl}C^{ijkl}={\cal R}_{ijkl}{\cal R}^{ijkl}-2 {\cal R}_{ij}{\cal R}^{ij}+\frac{1}{3}{\cal R}^2$ is the square of the Weyl tensor of $h_{ij}$.
In this equation, the left-hand side contains precisely the trace of the quasi-local energy-momentum tensor, $h_{ij}\frac{\delta S_{\rm ren}}{\delta h_{ij}} = -\frac{1}{2}\sqrt{-h}\, T^i{}_i$. Therefore we have found
\begin{equation}
T^i{}_i \,=\,\frac{1+4\lambda_2 \alpha g^2}{64\pi G g^3}\left({\cal R}_{ij}{\cal R}^{ij}-\frac{1}{3}{\cal R}^2-g^2 F^2\right) +\frac{\lambda_1\alpha}{16\pi G g}\left(C_{ijkl}C^{ijkl}-2g^2 F_{ij}F^{ij}\right)\,,
\end{equation}
where we also used \eqref{eq:Geff}. Pushing $\sqrt{-h}\, T^i{}_i$ to the conformal boundary gives a finite quantity, that is the trace of the holographic energy-momentum tensor.
This is to be compared with the general expression of the Weyl anomaly of ${\cal N}=1$ SCFTs,
\begin{equation}
T^{i}{}_i = -\frac{\aa}{16\pi^2} \hat{\cal X}_{\rm GB}+\frac{\cc}{16\pi^2}\left(\hat{C}_{ijkl}\hat{C}^{ijkl}-\frac{8}{3}{\hat F}_{ij}{\hat F}^{ij}\right)\,,
\end{equation}
where here $\hat{\cal X}_{\rm GB}$, $\hat{C}_{ijkl}\hat{C}^{ijkl}$ and ${\hat F}_{ij}{\hat F}^{ij}$ denote the Gauss-Bonnet, Weyl-squared and Maxwell terms constructed using the metric on the conformal boundary, and $\hat{F}_{ij}$ is the field strength of the gauge field $\hat A_i=\frac{\sqrt{3} g}{2} A_i$  coupling canonically to the R-current of the dual SCFT. The two expressions match upon using the dictionary \eqref{dictionary_ac} for the anomaly coefficients, that is what we wanted to verify.

\section{Conserved charges}\label{sec:conservedchargesgeneral}

The aim of this section is to derive conserved charges associated to the gauge symmetries of the theory: diffeomorphisms and U(1) gauge transformations. Before entering into the technical details of the derivation, let us briefly discuss the complications that one encounters when doing so. The main issue is that Noether currents associated to local symmetries are trivial, meaning that on-shell they can always be written as
\begin{equation}
J^{\mu}=\nabla_{\nu}k^{\mu\nu}\, ,
\end{equation}
where $k^{\mu\nu}=-k^{\nu\mu}$ so that the current is conserved. The main consequence of this is that the associated Noether charge is arbitrary. Let ${\cal C}$ be a Cauchy slice and integrate the Noether current on it:
\begin{equation}
Q =\int_{\cal C}{\boldsymbol \epsilon}_{\mu} J^{\mu}=\frac{1}{2}\int_{\partial {\cal C}}\,{\boldsymbol \epsilon}_{\mu\nu}\, k^{\mu\nu}\, ,
\end{equation}
where we have made use of the notation
\begin{equation}
{\boldsymbol\epsilon}_{\mu_1 \dots \mu_n}\equiv \frac{1}{\left(D-n\right)!}\epsilon_{\mu_1 \dots \mu_n \nu_1 \dots \nu_{D-n}}\, \diff x^{\nu_{1}}\wedge \dots \diff x^{\nu_{D-n}}\, ,
\label{def:boldsymbolepsilon}
\end{equation}
being $D$ the spacetime dimension. The charge defined in this way, though a conserved one, can take any value since at this stage $k^{\mu\nu}$ is arbitrary. Note that we could have defined another current,
\begin{equation}
{J'}^{\mu}=J^{\mu}+\nabla_{\nu}\left(\Delta k\right)^{\mu\nu}=\nabla_{\nu} \left(k^{\mu\nu}+\left(\Delta k\right)^{\mu\nu}\right)\,,
\end{equation}
which is also conserved and is therefore equivalent to $J^{\mu}$ if we do not ask for additional requirements. However, the associated charges do not coincide in general since
\begin{equation}
Q'-Q = \frac{1}{2}\int_{\partial{\cal C}}{\boldsymbol\epsilon}_{\mu\nu}\left(\Delta k\right)^{\mu\nu}\,.
\end{equation}

It turns out that the additional requirement that we have to impose in order to define the charge unambiguously is that the associated current vanishes on-shell, so that the $(D-2)$-form ${\bf k}=\frac{1}{2}{\boldsymbol \epsilon}_{\mu\nu}k^{\mu\nu}$ is conserved. 
This follows from the generalized Noether's theorem of Barnich, Brandt and Henneaux \cite{Barnich:2000zw}, which in simple terms states that there is a one-to-one correspondence between the equivalence classes of gauge parameters $\xi$ which are field symmetries (i.e., which leave  the fields invariant: $\delta_{\xi}\Psi=0$) and that of $(D-2)$-forms ${\mathbf k}_{\xi}$ which are conserved on-shell, $\diff{\mathbf k}_{\xi}=0$. We say that two gauge parameters are equivalent if their difference vanishes on-shell, whereas two conserved $(D-2)$-forms, ${\bf k}_{\xi}, {\bf k}'_{\xi}$ are equivalent if they differ by an exact form, ${\mathbf k'}_{\xi}\sim {\mathbf k}_{\xi}$ if ${\mathbf k'}_{\xi}={\mathbf k}_{\xi}+\diff{\mathbf l}_{\xi}$. 

\noindent
The charge is then given by 
\begin{equation}
Q_{\xi}=\int_{\partial {\cal C}} {\mathbf k}_{\xi}\,,
\end{equation}
and clearly does not depend on the representative of the equivalence class that is chosen to compute it since the integral of $\diff {\bf l}_{\xi}$ vanishes by Stokes theorem. For further details and for a pedagogic introduction to this topic we refer to~\cite{Compere:2006my, Compere:2018aar}.

\subsection{Wald formalism in the presence of Chern-Simons terms}\label{sec:Waldformalism}

In order to apply these ideas to the gauge symmetries of five-dimensional minimal gauged supergravity, we shall make use of the formalism developed by Wald in \cite{Wald:1993nt, Iyer:1994ys}, slightly modifying it to account for the effect of Chern-Simons terms. In particular, we will not demand the Lagrangian to be invariant under U$(1)$ gauge transformations, as it transforms by a total derivative when these terms are present -- see~\eqref{eq:delta_xiL2} below. As we will see, this crucially affects the definition and properties of the charges.

 Before getting started, let us introduce some notation that will be useful in the remaining of the paper. Following \cite{Wald:1993nt, Iyer:1994ys}, we define the $D$-form Lagrangian $\mathbf L$ as\footnote{In this section we are setting $16\pi G=1$ for convenience. It will be reinstated later.} 
\begin{equation}
\mathbf L={\boldsymbol\epsilon}\, {\mathcal L}\,, \hspace{1cm} S=\int {\bf L}\, .
\end{equation}
Under a generic variation of the fields, which we schematically denote by $\Psi=\{g_{\mu\nu}, A_{\mu}\}$, we have
\begin{equation}\label{eq:generaldL}
\delta {\mathbf L}={\mathbf E}_{\Psi}\delta \Psi+ \diff {\boldsymbol\Theta}(\Psi, \delta \Psi)\,,
\end{equation}
where ${\mathbf E}_{\Psi}=0$ are the equations of motion of the theory and $\boldsymbol\Theta(\Psi, \delta \Psi)$ is the term which is generated through integration by parts.

We now specify \eqref{eq:generaldL} to the most general gauge transformation given our field content. This consists of a diffeomorphism generated by a vector field $\xi^{\mu}$ plus a U$(1)$ gauge transformation parametrized by a function $\chi$, 
\begin{equation}\label{eq:deltafields}
\delta_{\xi}g_{\mu\nu}={\cal L}_{\xi}\, g_{\mu\nu}\,, \hspace{1cm} \delta_{\xi, \chi}A_{\mu}={\cal L}_{\xi} A_{\mu}+\nabla_{\mu}\chi\,,
\end{equation}
As in \cite{Elgood:2020svt}, we find it useful to introduce the so-called momentum map, $P_{\xi, \chi}$, which is defined as follows, 
\begin{equation}\label{eq:defP}
P_{\xi, \chi}=\chi+\iota_{\xi}A\, .
\end{equation}
Using it, we can write the transformation of the gauge field as
\begin{equation}\label{eq:delta_xiA}
\delta_{\xi, \chi} A=\iota_{\xi} F+ \diff P_{\xi, \chi}\,.
\end{equation}
When restricting to gauge parameters which are field symmetries, that is $\delta_{\xi}g=0$, $\delta_{\xi, \chi}A=0$, we have that $\xi$ becomes a Killing vector of the metric, as usual, while $P_{\xi, \chi}$ satisfies the momentum map equation
\begin{equation}\label{eq:momentummap}
\diff P_{\xi, \chi}=-\iota_{\xi}F\, ,
\end{equation}
which further implies $\delta_{\xi}F=0$.

The transformation \eqref{eq:deltafields} acts on the $D$-form Lagrangian $\bf L$ as follows
\begin{equation}\label{eq:delta_xiL1}
\begin{aligned}
\delta_{\xi, \chi}{\bf L}=\,&{\boldsymbol\epsilon}\left({\cal E}_{\mu\nu}\,\delta_{\xi}g^{\mu\nu}+{\cal E}^{\mu}\,\delta_{\xi, \chi}A_{\mu}\right)+\diff {\boldsymbol \Theta}_{\xi, \chi}\\[1mm]
=\,&{\boldsymbol\epsilon}\left[\nabla_{\mu}\left(-2\,\xi^{\nu}\,{\cal E}^{\mu}{}_{\nu}\,+P_{\xi, \chi}\,{\cal E}^{\mu}\right)\right]+\diff {\boldsymbol \Theta}_{\xi, \chi}\\[1mm]
=\,& \diff\left({\bf S}_{\xi, \chi}+{\boldsymbol\Theta}_{\xi, \chi}\right)\,,
\end{aligned}
\end{equation}
where we have defined 
\begin{equation}
{\boldsymbol\Theta}_{\xi, \chi}={\boldsymbol \Theta}\left(\Psi, \delta_{\xi, \chi}\Psi\right)\,, \hspace{1cm} {\bf S}_{\xi, \chi}={\boldsymbol\epsilon}_{\mu}\left(-2\,\xi^{\nu}\,{\cal E}^{\mu}{}_{\nu}\,+P_{\xi, \chi}\,{\cal E}^{\mu}\right)\,,
\end{equation}
and
\begin{equation}\label{def:eom}
{\cal E}_{\mu\nu}=\frac{\delta S}{e\, \delta g^{\mu\nu}}\,, \hspace{1cm} {\cal E}^{\mu}=\frac{\delta S}{e\, \delta A_{\mu}}\,.
\end{equation}
Moreover, we have used that (as a consequence of gauge invariance) ${\cal E}_{\mu\nu}$ and ${\cal E}^{\mu}$ satisfy the following off-shell identities,
\begin{equation}
2\nabla^{\mu}{\cal E}_{\mu\nu}=F_{\mu\nu}\,{\cal E}^{\mu}\, , \hspace{1cm} \nabla_{\mu}{\cal E}^{\mu}=0\,.
\end{equation}
In addition to \eqref{eq:delta_xiL1}, we assume that $\bf L$ transforms as follows,
\begin{equation}\label{eq:delta_xiL2}
\begin{aligned}
\delta_{\xi, \chi}{\bf L}\,=\,&{\cal L}_{\xi}{\bf L}+\diff\left(\chi{\boldsymbol \Lambda}\right)\,=\,\diff\left(\iota_{\xi}{\bf L}+\chi{\boldsymbol \Lambda}\right)\,,
\end{aligned}
\end{equation}
where $\boldsymbol\Lambda$ is a closed $(D-1)$-form that determines the transformation of ${\bf L}$ under gauge transformations. Hence, the results that we derive in this and the next section apply to theories whose Lagrangians satisfy Eq.~\eqref{eq:delta_xiL2}. A particular case is the Lagrangian of five-dimensional minimal gauged supergravity with four-derivative corrections presented in \eqref{eq:action5D}, which will be analyzed in detail in the next subsection. 
Putting together \eqref{eq:delta_xiL1} and \eqref{eq:delta_xiL2}, we obtain the conservation of the Noether current defined as
\begin{equation}\label{eq:Noethercurrent}
{\bf J}_{\xi, \chi}={\bf S}_{\xi, \chi}+{\boldsymbol \Theta}_{\xi, \chi}-\iota_\xi {\bf L}-\chi{\boldsymbol\Lambda}  \, ,
\end{equation}
which means that, at least locally,
\begin{equation}
{\bf J}_{\xi, \chi}=\diff {\bf Q}_{\xi, \chi}\,,
\end{equation}
being ${\bf Q}_{\xi, \chi}$ the Noether surface charge. Note that this is not yet the conserved $(D-2)$-form of the generalized Noether's theorem \cite{Barnich:2000zw}, as ${\bf J}_{\xi, \chi}$ does not vanish on-shell when $\xi, \chi$ are field symmetries. However, we can always improve it (adding a total derivative) in a way such that it does. Following the approach of \cite{Bazanski:1990qd, Kastor:2008xb, Ortin:2021ade}, we find that the improved current that vanishes on-shell is,
\begin{equation}\label{eq:improvedJ}
{\bf J}_{\xi, \chi}\to {\bf J}_{\xi, \chi}+\diff {\boldsymbol\Xi}_{\xi, \chi}\,, 
\end{equation}
where ${\boldsymbol\Xi}_{\xi, \chi}$ is a local $(D-2)$-form such that $\diff{\boldsymbol\Xi}_{\xi, \chi}=\iota_\xi {\bf L}+\chi{\boldsymbol\Lambda}$.\footnote{Notice that the integrability condition of ${\boldsymbol\Xi}_{\xi, \chi}$ is automatically satisfied if $\xi, \chi$ are field symmetries.} The surface charge ${\bf k}_{\xi, \chi}$ associated to this improved current  (and that is therefore conserved on-shell) is given by
\begin{equation}\label{eq:improvedcurrent}
{\bf k}_{\xi, \chi}={\bf Q}_{\xi, \chi}+{\boldsymbol\Xi}_{\xi, \chi}\,, \hspace{1cm} \diff{\bf k}_{\xi, \chi}={\bf J}_{\xi, \chi}+\diff {\boldsymbol\Xi}_{\xi, \chi}={\bf S}_{\xi, \chi}+{\boldsymbol \Theta}_{\xi, \chi}\, .
\end{equation}

Eventually, the charges that we obtain upon integration are,
\begin{equation}\label{eq:generalcharge}
{Q}_{\xi, \chi}=\int_{\partial\cal C}{\bf k}_{\xi, \chi}=\int_{\partial\cal C}\left({\bf Q}_{\xi, \chi}+{\boldsymbol\Xi}_{\xi, \chi}\right)\, .
\end{equation}
In the purely gravitational case, these charges are referred to as generalized Komar integrals, see e.g.~\cite{Komar:1958wp,Kastor:2009wy, Bazanski:1990qd, Kastor:2008xb, Ortin:2021ade} and references therein, while for a pure gauge symmetry they are called Page charges~\cite{Page:1983mke,Marolf:2000cb}. Their most interesting property  for the purposes of this work is that they obey the Gauss law. Let us assume that the spacetime can be foliated with a family of spacelike hypersurfaces $\Sigma_{z}$ diffeomorphic to the (conformal) boundary, labelled by a radial coordinate $z$. If regularity is assumed, then we have that
\begin{equation}\label{eq:Pageness}
\int_{\Sigma_{z}\, \cap \, {\cal C}}{\bf k}_{\xi, \chi}=\int_{\Sigma_{z'}\, \cap \, {\cal C}}{\bf k}_{\xi, \chi}\,,
\end{equation}
as a consequence of Stokes theorem. We will make use of this property to derive the quantum statistical relation in Section~\ref{sec:aladsthermodynamics} and to compute the corrected charges (electric charge and angular momentum) for the supersymmetric black hole of~\cite{Gutowski:2004ez} from the corrected near-horizon solution found in \cite{Cassani:2022lrk}. Let us then define the electric charge and angular momenta as particular cases of \eqref{eq:generalcharge}.

\textbf{Electric charge.} In order to define the electric charge, we restrict ourselves to a pure gauge transformation, i.e.~$\xi=0$. In this case, $\chi$ must be constant if it corresponds to a field symmetry and the improvement term ${\boldsymbol\Xi}_{0, \chi}$ simply reads
\begin{equation}\label{eq:chargexi}
{\boldsymbol\Xi}_{0, \chi}=\chi {\boldsymbol\Delta}\,, \hspace{1cm}\diff {\boldsymbol\Delta}={\boldsymbol\Lambda}\,,
\end{equation}
so that\footnote{We fix $\chi=1$ without loss of generality.} 
\begin{equation}\label{eq:pagecharge}
Q\equiv Q_{0, 1}=\int_{\partial {\cal C}}{\bf k}_{0, 1}=\int_{\partial {\cal C}}\left({\bf Q}_{0, 1}+{\boldsymbol\Delta} \right)\, .
\end{equation}
This corresponds with the notion of Page charge \cite{Page:1983mke,Marolf:2000cb}.

\textbf{Angular momenta.} Angular momenta can be defined as the Komar integral associated to the angular Killing vectors $\xi^{\mu}=\phi_{k}^{\mu}$, $k=1,2,\ldots$,
\begin{equation}\label{eq:defJi}
J_{k}=-\int_{\partial {\cal C}}{\bf k}_{\phi_k, 0}=-\int_{\partial {\cal C}}{\bf Q}_{\phi_k, 0}\, ,
\end{equation}
where we have used that the integral of the improvement term ${\boldsymbol\Xi}_{\xi, 0}$ vanishes when $\xi^{t}=0$.

\subsection{Specifying to five-dimensional higher-derivative supergravity}\label{sec:conservedchargesourtheory}

Let us now specify to the four-derivative theory of interest to us \eqref{eq:action5D} and derive the corresponding expressions for the Noether current \eqref{eq:Noethercurrent} and surface charge \eqref{eq:generalcharge}. Let us start defining the following auxiliary tensors,\footnote{We note that  ${\cal P}^{\mu\nu\rho\sigma}={P}^{\mu\nu\rho\sigma}+{\Pi}^{\mu\nu\rho\sigma}+ \text{(anti)symmetrizations}$, where $\displaystyle P^{\mu\nu\rho\sigma}=\frac{\partial {\cal L}'}{\partial R_{\mu\nu\rho\sigma}}$ is the tensor used in Appendix C of \cite{Cassani:2022lrk} in order to write down the equations of motion. The explicit expression for $\displaystyle \frac{\partial {\cal L}'}{\partial F_{\mu\nu}}$ can also be found there.}
\begin{equation}\label{eq:auxiliarytensors}
{\cal P}^{\mu\nu\rho\sigma} = \frac{\partial\mathcal L}{\partial R_{\mu\nu\rho\sigma}}\,\,,\hspace{1cm} {\mathcal F}^{\mu\nu} =-2\,\frac{\partial\mathcal L'}{\partial F_{\mu\nu}}\,\,, \hspace{1cm}\Pi^{\mu\nu\rho\sigma} = -\frac{\lambda_1\alpha}{\sqrt{3}}\epsilon^{\mu\nu\alpha\beta\gamma}R_{\alpha\beta}{}^{\rho \sigma}A_{\gamma}\,\,,\hspace{1cm}
\end{equation}
where ${\mathcal L'}={\cal L}-{\cal L}_{\rm{CS}}$ is the Lagrangian density without the contribution of the Chern-Simons terms, which are treated separately. It is assumed (and there is no loss of generality in doing so) that the above auxilary tensors inherit the algebraic symmetries of the associated curvature tensors, namely
\begin{equation}
\mathcal P^{(\mu\nu)\rho\sigma} =\, 0 \,\,, \hspace{7.5mm} {\mathcal P}^{\mu\nu\rho\sigma}={\mathcal P}^{\rho\sigma\mu\nu}\,\,,\hspace{7.5mm}{\mathcal P}^{\mu[\nu\rho\sigma]}=0\,\,,\hspace{7.5mm}{\mathcal F}^{(\mu\nu)} = 0\,\,.
\label{eq:propertiesauxiliarytensors}
\end{equation}
According to \eqref{eq:Noethercurrent}, the information we need in order to compute the Noether current ${\bf J}_{\xi, \chi}$ consists of the tensors ${\cal E}_{\mu\nu}$, ${\cal E}^{\mu}$, the boundary term ${\boldsymbol\Theta}={\boldsymbol \epsilon}_{\mu}\Theta^{\mu}$ and $\boldsymbol \Lambda={\boldsymbol \epsilon}_{\mu}\Lambda^{\mu}$. Making use of the results contained in Appendix~C of \cite{Cassani:2022lrk} and of the definitions \eqref{eq:auxiliarytensors}, we have that
\begin{equation}\label{eq:eoms2}
\begin{aligned}
\mathcal E_{\mu\nu}=\,&\mathcal P_{(\mu}{}^{\rho\sigma\lambda} R_{\nu)\rho\sigma\lambda} - 2\nabla^\rho\nabla^\sigma\mathcal P_{\rho(\mu\nu)\sigma} +\frac{1}{2} {\mathcal F}_{\rho(\mu}F_{\nu)}^{\,\,\,\,\rho} -\frac{1}{2}g_{\mu\nu} \mathcal L' -\Pi_{(\mu}^{\,\,\,\,\,\rho\sigma\lambda} R_{\nu)\rho\sigma\lambda}\,\,,\\[1mm]
\mathcal E^\mu =\,& \nabla_\nu\mathcal F^{\nu\mu} - \frac{c_3}{4\sqrt{3}}\epsilon^{\mu\nu\rho\sigma\lambda} F_{\nu\rho} F_{\sigma\lambda} - \frac{\lambda_1\alpha}{2\sqrt{3}}\epsilon^{\mu\nu\rho\sigma\lambda} R^{\delta\gamma}_{\,\,\,\,\,\,\,\nu\rho} R_{\delta\gamma\sigma\lambda}\,\,,
\end{aligned}
\end{equation}
and
\begin{equation}\label{eq:theta}
{\Theta}^{\mu} = 2\mathcal P^{\rho\sigma\mu\nu}\nabla_\sigma\delta g_{\nu\rho} - 2\nabla_\sigma\mathcal P^{\mu\nu\rho\sigma} \delta g_{\nu\rho} -{\mathcal F}^{\mu\nu} \delta A_\nu - \frac{c_3}{3\sqrt{3}}\epsilon^{\mu\nu\rho\sigma\lambda} \delta A_\nu A_\rho F_{\sigma\lambda}\,\,.
\end{equation}
It will be useful later to split the boundary term ${\boldsymbol\Theta}$ into its ``gravitational'' and ``electromagnetic'' contributions, ${\boldsymbol\Theta}={\boldsymbol\Theta}^{\rm g}+{\boldsymbol\Theta}^{\rm {em}}$, where
\begin{equation}\label{eq:split_theta}
\begin{aligned}
{\boldsymbol\Theta}^{\rm g}=\,&{\boldsymbol \epsilon}_{\mu}\left(2\mathcal P^{\rho\sigma\mu\nu}\nabla_\sigma\delta g_{\nu\rho} - 2\nabla_\sigma\mathcal P^{\mu\nu\rho\sigma} \delta g_{\nu\rho}\right)\,,\\[1mm]
{\boldsymbol\Theta}^{\rm em}=\,&{\boldsymbol \epsilon}_{\mu}\left( -{\mathcal F}^{\mu\nu} \delta A_\nu - \frac{c_3}{3\sqrt{3}}\epsilon^{\mu\nu\rho\sigma\lambda} \delta A_\nu A_\rho F_{\sigma\lambda}\right)\,.
\end{aligned}
\end{equation}
Finally, the transformation of the action under U$(1)$ gauge transformations is characterized by 
\begin{equation}
{\Lambda}^\mu =\,-\epsilon^{\mu\nu\rho\sigma\lambda} \left(\frac{c_3}{12\sqrt{3}} F_{\nu\rho} F_{\sigma\lambda} + \frac{\lambda_1\alpha}{2\sqrt{3}}R_{\nu\rho\delta\gamma} R_{\sigma\lambda}{}^{\delta\gamma}\right)\,, \hspace{1cm} \Lambda^{\mu}=\nabla_{\nu}\Delta^{\mu\nu}\,,
\label{eq:lambda}
\end{equation}
where
\begin{equation}\label{eq:Deltamunu}
\Delta^{\mu\nu} = -\epsilon^{\mu\nu\rho\sigma\lambda}\left[\frac{c_3}{6\sqrt{3}}A_\rho F_{\sigma\lambda} +\frac{\lambda_1\alpha}{3\sqrt{3}}\Omega_{\rm{CS}}{}_{\rho\sigma\lambda}\right]\,\,,
\end{equation}
and $\Omega_{\rm{CS}}$ is the Lorentz-Chern-Simons three-form constructed out of the gravitational spin connection $\omega^{ab}=\,\omega_{\mu}^{ab}\,\diff x^{\mu}$,
\begin{equation}\label{eq:lorentzchernsimons}
\Omega_{\rm{CS}}=\diff \omega^{ab}\wedge \omega_{ab}-\frac{2}{3}\, \omega^{a}{}_{b}\wedge \omega^{b}{}_{c}\wedge \omega^{c}{}_a\,,
\end{equation}
where $a, b, \dots$ are flat indices.

Putting these results together, we find the following expression for the Noether current ${\bf J}_{\xi, \chi}={\boldsymbol \epsilon}_{\mu}J^\mu_{\xi,\chi}$, 
\begin{equation}
\begin{aligned}
J^\mu_{\xi,\chi}=\, &\left(-2\mathcal P^{(\mu|\rho\sigma\lambda} R^{|\nu)}_{\,\,\,\,\,\rho\sigma\lambda} +4\nabla_\rho\nabla_\sigma\mathcal P^{\rho(\mu\nu)\sigma} - \mathcal F^{\rho(\mu}F^{\nu)}_{\,\,\,\,\rho}\right)\xi_\nu +P_{\xi,\chi}\nabla_\nu\mathcal F^{\nu\mu}\\[1mm]
+& 2\mathcal P^{\rho\sigma\mu\nu}\nabla_\sigma\delta_{\xi} g_{\nu\rho} - 2\nabla_\sigma\mathcal P^{\mu\nu\rho\sigma} \delta_{\xi} g_{\nu\rho} -\mathcal F^{\mu\nu} (\delta_{\xi,\chi} A)_\nu - \xi^\mu \mathcal L_{\rm{CS}}\\[1mm]
-& P_{\xi,\chi}\epsilon^{\mu\nu\rho\sigma\lambda}\left(\frac{c_3}{4\sqrt{3}} F_{\nu\rho} F_{\sigma\lambda} +\frac{\alpha\lambda_1}{2\sqrt{3}} R^{\delta\gamma}{}_{\nu\rho} R_{\sigma\lambda\delta\gamma}\right) - \frac{c_3}{3\sqrt{3}}\epsilon^{\mu\nu\rho\sigma\lambda} (\delta_{\xi,\chi} A)_\nu A_\rho F_{\sigma\lambda}-\\[1mm]
&-\frac{2\lambda_1\alpha}{\sqrt{3}}\epsilon^{(\mu|\rho\alpha\beta\gamma}A_\alpha R_{\beta\gamma}{}^{\sigma\lambda}R^{|\nu)}{}_{\rho\sigma\lambda} \xi_\nu +\chi\epsilon^{\mu\nu\rho\sigma\lambda}  \left(\frac{c_3}{12\sqrt{3}}F_{\nu\rho} F_{\sigma\lambda} +\frac{\alpha\lambda_1}{2\sqrt{3}}R_{\nu\rho}{}^{\delta\gamma} R_{\sigma\lambda\delta\gamma}\right).
\end{aligned}
\label{eq:oneformnoethercurrent}\end{equation}
Using the transformation rules \eqref{eq:deltafields} and the fact that
\begin{equation}
\begin{aligned}
-\xi^\mu\mathcal L_{\rm{CS}} &\,=\,\epsilon^{\mu\nu\rho\sigma\lambda}\left[\iota_\xi A  \left(\frac{c_3}{12\sqrt{3}}F_{\nu\rho} F_{\sigma\lambda} +\frac{\alpha\lambda_1}{2\sqrt{3}}R^{\delta\gamma}_{\,\,\,\,\,\,\,\nu\rho} R_{\delta\gamma\sigma\lambda}\right) \right.\\[1mm]
&\left.\qquad\quad\  +\,\xi^\tau A_\rho\left(\frac{c_3}{3\sqrt{3}}F_{\tau\nu} F_{\sigma\lambda}
 +\frac{2\alpha\lambda_1}{\sqrt{3}}R_{\tau\nu\delta\gamma} R_{\sigma\lambda}^{\,\,\,\,\,\,\,\delta\gamma} \right)\right],
\end{aligned}
\label{eq:chernsimonsidentity}\end{equation}
it is possible to manipulate \eqref{eq:oneformnoethercurrent}, extending the calculation in Appendix A of~\cite{Ortin:2021ade} to our more general setup, to obtain that
\begin{equation}\label{eq:oneformnoethercurrent2}
{\bf J}_{\xi,\chi} =\, {\boldsymbol \epsilon}_{\mu}\nabla_\nu\left[-4\nabla_\sigma \mathcal P^{\nu\mu\sigma\rho}\xi_\rho +2 \mathcal P^{\nu\mu\sigma\rho}\nabla_\sigma\xi_\rho +P_{\xi,\chi}\left(\mathcal F^{\nu\mu} +\frac{c_3}{3\sqrt{3}}\epsilon^{\nu\mu\rho\sigma\lambda}A_\rho F_{\sigma\lambda} \right)\right]\,.
\end{equation}
The associated Noether surface charge ${\bf Q}_{\xi, \chi}$ is therefore given by \begin{equation}\label{eq:noethersurfacechargecomponents}
{\bf Q}_{\xi,\chi} =\,\frac{1}{2}\boldsymbol {\epsilon}_{\mu\nu}\left[4\nabla_\sigma \mathcal P^{\mu\nu\sigma\rho}\xi_\rho - 2 \mathcal P^{\mu\nu\sigma\rho}\nabla_\sigma\xi_\rho - P_{\xi,\chi}\left( \mathcal F^{\mu\nu} +\frac{c_3}{3\sqrt{3}}\epsilon^{\mu\nu\rho\sigma\lambda}A_\rho F_{\sigma\lambda} \right)\right]\,,
\end{equation}
and we observe that it can be written as the sum of two contributions,
\begin{equation}\label{eq:surfacecharge2}
\mathbf Q_{\xi,\chi} = \mathbf Q_\xi^{\rm g} +P_{\xi,\chi} \,\mathbf Q_{0,1}\,\,,
\end{equation}
where ${\bf Q}^{\rm g}_{\xi}$ is the generalized Komar $(D-2)$-form \cite{Compere:2006my},
\begin{equation}
{\bf Q}^{\rm g}_{\xi} = \,\frac{1}{2}\boldsymbol {\epsilon}_{\mu\nu}\left(4\nabla_\sigma \mathcal P^{\mu\nu\sigma\rho}\xi_\rho - 2 \mathcal P^{\mu\nu\sigma\rho}\nabla_\sigma\xi_\rho\right)\,\,, 
\label{eq:defQg}
\end{equation}
and 
\begin{equation}\label{eq:Q_{0,1}}
{\bf Q}_{0,1} = -\frac{1}{2}\boldsymbol {\epsilon}_{\mu\nu}\left(\mathcal F^{\mu\nu} +\frac{c_3}{3\sqrt{3}}\epsilon^{\mu\nu\rho\sigma\lambda}A_\rho F_{\sigma\lambda} \right)\,\,.
\end{equation}

Let us provide an explicit expression for the electric charge defined in \eqref{eq:pagecharge}. Taking into account \eqref{eq:Q_{0,1}} and the expression for ${\boldsymbol\Delta}$ given in \eqref{eq:Deltamunu}, we find that 
\begin{equation}\label{eq:k_{0,1}}
\begin{aligned}
{\bf k}_{0,1} =\,& -\frac{1}{2}\boldsymbol \epsilon_{\mu\nu}  \left[ \mathcal F^{\mu\nu} +\epsilon^{\mu\nu\rho\sigma\lambda}\left(\frac{c_3}{2\sqrt{3}}A_\rho F_{\sigma\lambda} + \frac{\lambda_1\alpha}{3\sqrt{3}}\Omega_{\rm{CS}}{}_{\rho\sigma\lambda}\right)\right]\,\,,
\end{aligned}
\end{equation}
and therefore the electric charge reduces to
\begin{equation}
\begin{aligned}
Q=-\int_{\partial {\cal C}}\left(\star {\cal F}-\frac{c_3}{\sqrt{3}}F\wedge A-\frac{2\lambda_1\alpha}{\sqrt{3}}\Omega_{\rm{CS}}\right)\,,
\end{aligned}
\end{equation}
which coincides with the notion of Page charge, as anticipated. 

Finally, we can give the expression for the angular momentum \eqref{eq:defJi}. Considering the surface charge \eqref{eq:noethersurfacechargecomponents} for an angular Killing vector $\phi_k$ and taking $\chi=0$ gives
\begin{equation}\label{eq:surfacechargeangmom}
 J_{k} = -\frac{1}{2}\int_{\partial {\cal C}}\!\boldsymbol {\epsilon}_{\mu\nu}\left[4\nabla_\sigma \mathcal P^{\mu\nu\sigma\rho}(\phi_k)_\rho - 2 \mathcal P^{\mu\nu\sigma\rho}\nabla_\sigma(\phi_k)_\rho - \left(\iota_{\phi_k} A\right)\left( \mathcal F^{\mu\nu} +\frac{c_3}{3\sqrt{3}}\epsilon^{\mu\nu\rho\sigma\lambda}A_\rho F_{\sigma\lambda} \right)\right].
\end{equation}

In particular, we can apply the above integral to evaluate the angular momentum for the two-derivative minimal gauged supergravity, which is obtained by taking $\alpha= 0$ in the action \eqref{eq:action5D}. In this case, the explicit expressions for the tensors \eqref{eq:auxiliarytensors} are particularly simple,
\begin{equation}
\mathcal P^{\mu\nu\rho\sigma} = g^{\mu[\rho}g^{\sigma]\nu}\,\,,\quad\quad \mathcal F^{\mu\nu} = F^{\mu\nu}\,\,,
\end{equation}
and the angular momentum \eqref{eq:surfacechargeangmom} reproduces the one given in~\cite{Barnich:2005kq,Suryanarayana:2007rk,Hanaki:2007mb,Cassani:2018mlh}.

\section{Black hole thermodynamics}\label{sec:aladsthermodynamics}

In this section we consider stationary spacetimes with an event horizon and derive the first law of black hole mechanics \cite{Bardeen:1973gs} and the quantum statistical relation \cite{Gibbons:1976ue} in the class of theories for which the results of the previous section apply.  For these purposes, we will follow Wald's approach~\cite{Wald:1993nt}, suitably taking into account the contribution of the gauge field. The strategy will be to define a conserved charge which takes the same value when integrated at the horizon and at infinity, and then identify the various contributions in terms of the conserved charges and entropy of the solution.

The first law in the Einstein-Maxwell theory has been extensively studied in the past making use of Wald's formalism, see e.g.~\cite{Gao:2001ut, Gao:2003ys, Compere:2006my, Compere:2007vx, Prabhu:2015vua,  Elgood:2020svt, Hajian:2022lgy} and references therein. Depending on the gauge choice one makes, the contribution of the gauge field can arise either from the integral at the horizon or from the integral at infinity (or from both if one works in a general gauge). In all cases one obtains exactly the same term, $-\Phi\,\delta Q$, with the gauge choice just manifesting in the identification of the electrostatic chemical potential $\Phi$. The reason behind is that the Noether charge associated to U$(1)$ transformations in the Einstein-Maxwell theory, $Q\sim\int \star F$,  is gauge invariant and satisfies Gauss law, which implies that the charges computed at the horizon and at infinity coincide.\footnote{This naturally extends to theories whose Lagrangians only depend on the gauge field $A_{\mu}$ through its field strength $F_{\mu\nu}$.} As we have seen in the previous section, in presence of Chern-Simons terms the Noether charge is neither gauge invariant nor it satisfies the Gauss law. Indeed, defining a charge obeying the Gauss law was our main motivation for introducing the Page charge \eqref{eq:pagecharge}, which in any case fails to be gauge invariant as well. This raises the question of which charge appears in the first law (and quantum statistical relation). To the best of our knowledge, this is still an open question.
 Our strategy to deal with this issue will be the following. First, we will set ourselves in the regular gauge, which means that we will impose 
\begin{equation}\label{eq:regulargauge}
\iota_{\xi}A|_{\cal H}=0\,,
\end{equation}
where $\xi=t+\Omega_k\, \phi_k$ is the generator of the horizon, with $t$ the Killing vector generating time translations, $\Omega_k$ the angular velocity, and ${\cal H}$ is an arbitrary slice of the horizon. In this gauge, the contributions of the gauge field appear at infinity, where we further assume that
\begin{equation}\label{eq:assumptions}
\iota_{\phi_k}A|_{\partial{\cal M}\, \cap \, {\cal C}}=0\, , \hspace{1cm}{\boldsymbol \Delta}|_{\partial{\cal M}\, \cap \, {\cal C}}=0\, .
\end{equation}
Crucially, this implies that the Page charge \eqref{eq:pagecharge} matches the Noether charge computed asymptotically, namely
\begin{equation}\label{eq:equiv_charges}
Q=\int_{{\partial\cal M}\, \cap \, {\cal C}}{\bf Q}_{0, 1}\, ,
\end{equation}
and also that the angular momenta \eqref{eq:defJi} can be computed at infinity by the following integral,
\begin{equation}\label{eq:Ji_inf}
J_{k}=-\int_{\partial {\cal M}\, \cap \, {\cal C}}{\bf Q}^{\rm g}_{\phi_k}\, .
\end{equation}
Let us notice that the condition \eqref{eq:assumptions} holds for any solution that is asymptotically AdS, as opposed to asymptotically {\it locally} AdS. 
%On the other hand, in most cases it is possible to perform a gauge transformation along the angular directions such that \eqref{eq:assumptions} is verified, but for AlAdS solutions one should check case by case that the transformation does not lead to singularities in the solution.

Eventually, we will show that the asymptotic integrals \eqref{eq:equiv_charges} and \eqref{eq:Ji_inf}, where the notions of Page, Komar and Noether charges coincide, are the ones that enter the first law and the quantum statistical relation.

In what follows, we specify $\xi$ to be the generator of the horizon and take $\chi$ to be constant, which trivializes the gauge transformations. In particular, one can take $\chi=0$ without loss of generality.

\subsection{First law of black hole mechanics}\label{sec:firstlaw}

The variation of the Noether current \eqref{eq:Noethercurrent} under a generic perturbation of the fields (keeping $\xi$ and $\chi$ fixed) is given by,
\begin{equation}
\delta {\bf J}_{\xi, \chi}=\delta {\bf S}_{\xi, \chi}+\delta {\boldsymbol \Theta_{\xi, \chi}}-\iota_{\xi}\delta {\bf L}-\chi \delta {\boldsymbol \Lambda}\, .
\end{equation}
Restricting to perturbations obeying the linearized equations of motion and going on-shell, one finds
\begin{equation}
\begin{aligned}
\delta {\bf J}_{\xi, \chi}=\,&\delta {\boldsymbol \Theta_{\xi, \chi}}-\iota_{\xi}\diff {\boldsymbol \Theta}-\chi \delta {\boldsymbol \Lambda}\\[1mm]
=\,&\delta {\boldsymbol \Theta_{\xi, \chi}}-\delta_{\xi, \chi} {\boldsymbol \Theta}+\diff\left(\iota_{\xi}{\boldsymbol \Theta}-\chi \delta {\boldsymbol \Delta}\right)\\[1mm]
=\,& \diff \left(\delta {\bf Q}_{\xi, \chi}\right)\, ,
\end{aligned}
\end{equation}
so that
\begin{equation}
\diff\left(\delta {\bf Q}_{\xi, \chi}-\iota_{\xi}{\boldsymbol \Theta}+\chi \delta {\boldsymbol \Delta}\right)=\delta {\boldsymbol \Theta_{\xi, \chi}}-\delta_{\xi, \chi} {\boldsymbol \Theta}\, .
\end{equation}
When further restricting to field symmetries, the right-hand side vanishes, which means that the surface charge,
\begin{equation}\label{eq:surfacecharge1stlaw}
\delta {\bf Q}_{\xi, \chi}-\iota_{\xi}{\boldsymbol \Theta}+\chi \delta {\boldsymbol \Delta}\,, 
\end{equation}
is conserved on-shell. 

Let $\cal C$ be a Cauchy slice and denote as $\cal H$ its intersection with the event horizon. In particular, one can think of $\cal H$ as the bifurcation surface, as originally assumed by Wald \cite{Wald:1993nt}. However, according to \cite{Jacobson:1993vj}, the results should extend to any slice of the horizon, as long as regularity at the bifurcation surface is guaranteed. Assuming regularity of the fields on and outside the horizon so that it is allowed to apply Stokes theorem, we have that
\begin{equation}\label{eq:horizonintegral}
\int_{\partial {\cal M}\, \cap \, {\cal C}}\left(\delta {\bf Q}_{\xi, \chi}-\iota_{\xi}{\boldsymbol \Theta}+\chi \delta {\boldsymbol \Delta}\right)=\int_{\cal H}\left(\delta {\bf Q}_{\xi, \chi}-\iota_{\xi}{\boldsymbol \Theta}+\chi \delta {\boldsymbol \Delta}\right)\,,
\end{equation}
from which stems the first law, as we shall see. Setting $\chi=0$ and splitting the asymptotic contributions into the gravitational and electromagnetic pieces using \eqref{eq:surfacecharge2} and \eqref{eq:split_theta}, we have 
\begin{equation}
\int_{\partial {\cal M}\, \cap \, {\cal C}}\left(\delta {\bf Q}_{\xi, 0}-\iota_{\xi}{\boldsymbol \Theta}\right)=\, \int_{\partial {\cal M}\, \cap \, {\cal C}}\left(\delta{\bf Q}^{\rm g}_{\xi}-\iota_{\xi}{\boldsymbol \Theta}^{\rm g}\right)+\int_{\partial {\cal M}\, \cap \, {\cal C}}\left(\iota_{\xi}\delta A \,{\bf Q}_{0, 1}+\iota_{\xi}A \,\delta{\bf Q}_{0, 1}-\iota_{\xi}{\boldsymbol \Theta}^{\rm{em}}\right)\,.
\end{equation}
The first term yields \cite{Wald:1993nt, Iyer:1994ys},
\begin{equation}
\int_{\partial {\cal M}\, \cap \, {\cal C}}\left(\delta{\bf Q}^{\rm g}_{\xi}-\iota_{\xi}{\boldsymbol \Theta}^{\rm g}\right)=\delta E-\Omega_k\, \delta J_{k}\, .
\end{equation}
Instead the second term gives us the contribution of the electric charge to the first law, which by our gauge choice appears from the integral at infinity, as already anticipated. In order to see this explicitly, let us massage it as follows 
\begin{equation}
\begin{aligned}
\int_{\partial {\cal M}\, \cap \, {\cal C}}\left(\iota_{\xi}\delta A \,{\bf Q}_{0, 1}+\iota_{\xi}A \,\delta{\bf Q}_{0, 1}-\iota_{\xi}{\boldsymbol \Theta}^{\rm {em}}\right)=\,& \int_{\partial {\cal M}\, \cap \, {\cal C}}\left(-\Phi \,\delta {\bf Q}_{0, 1}+\iota_{\xi}{\bf Q}_{0, 1}\wedge \delta A\right)\\[1mm]
=&\int_{\partial {\cal M}\, \cap \, {\cal C}}\left(-\Phi \,\delta {\bf Q}_{0, 1}+\iota_{\xi}{\bf k}_{0, 1}\wedge \delta A-\iota_{\xi}{\boldsymbol\Delta}\wedge \delta A\right)\,.
\end{aligned}
\end{equation}
Now we notice that the second and third term in the second line do not contribute, provided the assumptions \eqref{eq:assumptions} hold.\footnote{The second term has been recently understood in \cite{Ortin:2022uxa, Ballesteros:2023iqb} to give rise (after integration by parts) to a term $\sim-{\Phi}_{\rm m}\,{\delta Q}_{\rm{m}}$, where $Q_{\rm m}$ are the magnetic charges and $\Phi_{\rm m}$ the associated chemical potentials. However, our initial assumptions \eqref{eq:assumptions} forbid this kind of term.} In such case, we obtain
\begin{equation}
\int_{\partial {\cal M}\, \cap \, {\cal C}}\left(\iota_{\xi}\delta A \,{\bf Q}_{0, 1}+\iota_{\xi}A \,\delta{\bf Q}_{0, 1}-\iota_{\xi}{\boldsymbol \Theta}^{\rm {em}}\right)=-\Phi\, \delta\int_{\partial {\cal M}\, \cap \, {\cal C}}{\bf Q}_{0, 1}=-\Phi \,\delta Q\,,
\end{equation}
where we have made use of \eqref{eq:equiv_charges}. Putting together all the asymptotic contributions,

\begin{equation}\label{eq:1stlaw_inf}
\int_{\partial {\cal M}\, \cap \, {\cal C}}\left(\delta {\bf Q}_{\xi, 0}-\iota_{\xi}{\boldsymbol \Theta}\right)=\delta E-\Omega_k \,\delta J_k -\Phi \,\delta Q\, .
\end{equation}
If the first law holds in its standard form, we should be able to identify the horizon contribution with $\beta^{-1}\delta {\cal S}$, where $\beta$ is the inverse Hawking temperature and ${\cal S}$ is the black hole entropy \cite{Wald:1993nt}. Indeed, when regularity of the fields at the bifurcation surface is assumed, the horizon integral reduces to \cite{Wald:1993nt, Iyer:1994ys, Jacobson:1993vj},
\begin{equation}
\int_{\cal H}\left(\delta {\bf Q}_{\xi, 0}-\iota_{\xi}{\boldsymbol \Theta}\right)\,=\,\delta\int_{\cal H}{\bf Q}^{\rm g}_{\xi}\,=\,\beta^{-1}\delta \left(-2\pi\int_{\cal H} \diff^3x\, \sqrt{\gamma}\,{\cal P}^{\mu\nu\rho\sigma}\, n_{\mu\nu}\, n_{\rho\sigma}\right)\, ,
\end{equation}
where $\gamma$ is the metric induced at ${\cal H}$ and $n_{\mu\nu}$ the binormal normalized so that $n_{\mu\nu}n^{\mu\nu}=-2$. This allows us to infer the expression for the black hole entropy,
\begin{equation}\label{eq:Waldentropy}
{\cal S}=-2\pi\int_{\cal H} \diff^3x\, \sqrt{\gamma}\,{\cal P}^{\mu\nu\rho\sigma}\, n_{\mu\nu}\, n_{\rho\sigma}\,,
\end{equation}
which coincides with Wald's prescription \cite{Wald:1993nt}. This formula was already tested in \cite{Cassani:2022lrk}, where it was applied to compute the corrected entropy of supersymmetric AdS$_5$ black holes, obtaining a perfect agreement with Euclidean methods.

\subsection{Quantum statistical relation} \label{sec:qsr}

In order to derive the quantum statistical relation \cite{Gibbons:1976ue}, we start from the equality between the integral of the surface charge ${\bf k}_{\xi, \chi}$ at infinity and at the horizon,
\begin{equation}\label{eq:raw-qsr}
\int_{\Sigma_{\epsilon} \,\cap \,{\cal C}}{\bf k}_{\xi, \chi}=\int_{\cal H}{\bf k}_{\xi, \chi}\,.
\end{equation}
We recall that $\xi=t+\Omega_k\, \phi_k$ is the generator at the horizon and $\chi$ is assumed to be constant. Making use of \eqref{eq:improvedcurrent}, we find
\begin{equation}
\begin{aligned}
\int_{\Sigma_{\epsilon} \,\cap \,{\cal C}}{\bf Q}_{\xi, \chi}-\int_{\cal H}{\bf Q}_{\xi, \chi}=\,&-\int_{\Sigma_{\epsilon} \,\cap \,{\cal C}}{\boldsymbol \Xi}_{\xi, \chi}+\int_{\cal H}{\boldsymbol \Xi}_{\xi, \chi}\\[1mm]
=\,&-\int_{\cal C}\diff{\boldsymbol \Xi}_{\xi, \chi}=-\int_{\cal C} \left[\iota_{\xi}{\bf L}+\diff\left(\chi{\boldsymbol \Delta}\right)\right]\,.
\end{aligned}
\end{equation}
Next, we add suitable boundary terms,
\begin{equation}\label{eq:QSR}
\int_{\Sigma_{\epsilon} \,\cap \,{\cal C}}\left({\bf Q}_{\xi, \chi}+\iota_{\xi}{\bf B}+\chi {\boldsymbol \Delta}\right)-\int_{\cal H}\left({\bf Q}_{\xi, \chi}+\chi {\boldsymbol \Delta}\right)=-\int_{\cal C}\iota_{\xi}{\bf L}+\int_{\Sigma_{\epsilon} \,\cap \,{\cal C}} \iota_{\xi}{\bf B}\,,
\end{equation}
so that we can identify the right-hand side with $\beta^{-1}I$ (where $I$ is the renormalized Euclidean on-shell action) when the cutoff is removed~\cite{Papadimitriou:2005ii}. Then, 
\begin{equation}\label{eq:qsr2}
\beta^{-1}I=\int_{\partial {\cal M} \,\cap \,{\cal C}}\left({\bf Q}_{\xi, \chi}+\iota_{\xi}{\bf B}+\chi {\boldsymbol \Delta}\right)-\int_{\cal H}\left({\bf Q}_{\xi, \chi}+\chi {\boldsymbol \Delta}\right)\, .
\end{equation}

Let us first consider the integral at infinity. Setting $\chi=0$ and using  \eqref{eq:surfacecharge2} again, we have
\begin{equation}
\int_{\partial {\cal M} \,\cap \,{\cal C}}\left({\bf Q}_{\xi, 0}+\iota_{\xi}{\bf B}\right)=\,\int_{\partial {\cal M} \,\cap \,{\cal C}}\left({\bf Q}^{\rm g}_{t}+\iota_{t}{\bf B}\right)+\Omega_k\int_{\partial {\cal M} \,\cap \,{\cal C}}{\bf Q}^{\rm g}_{\phi_k}-\Phi\int_{\partial {\cal M} \,\cap \,{\cal C}}{\bf Q}_{0, 1}\,.
\end{equation}
As discussed in \cite{Wald:1993nt, Papadimitriou:2005ii}, the first term can be interpreted as the mass of the solution,
\begin{equation}\label{eq:mass}
E=\int_{\partial {\cal M} \,\cap \,{\cal C}}\left({\bf Q}^{\rm g}_{t}+\iota_{t}{\bf B}\right)\,,
\end{equation}
while the second and third yield (minus) the angular momenta and Page charge, as we showed in and \eqref{eq:Ji_inf} and \eqref{eq:equiv_charges}. Hence,
\begin{equation}\label{eq:qsr_inf}
\int_{\partial {\cal M} \,\cap \,{\cal C}}\left({\bf Q}_{\xi, 0}+\iota_{\xi}{\bf B}\right)=\,E-\Omega_{k} J_{k}-\Phi \,Q\, .
\end{equation}
On the other hand, the horizon contribution in \eqref{eq:qsr2} boils down to
\begin{equation}\label{eq:qsr_hor}
\int_{\cal H}{\bf Q}_{\xi, 0}=\int_{\cal H}{\bf Q}^{\rm g}_{\xi}=\beta^{-1}{\cal S}\, ,
\end{equation}
where ${\cal S}$ is given in \eqref{eq:Waldentropy}. Finally, plugging \eqref{eq:qsr_inf} and \eqref{eq:qsr_hor} into \eqref{eq:qsr2}, we get the quantum statistical relation
\begin{equation}\label{eq:qsr}
\beta^{-1}I=E-\beta^{-1}{\cal S}-\Omega_k\, J_k-\Phi\, Q\,,
\end{equation}
as we wanted to show.

\section{Corrected charges of BPS AdS$_5$ black hole from near-horizon geometry}\label{sec:corrections_GR}

In this section we apply the formalism above to the supersymmetric black hole solution to five-dimensional minimal gauged supergravity. In particular, we want to reproduce the corrected thermodynamical charges directly from the near-horizon geometry. We expect to be able to do that since in Section \ref{sec:conservedchargesgeneral} we have constructed charges satisfying a Gauss law, that can be equivalently evaluated on any radial slice of the spacetime at fixed time.\footnote{The same approach has been followed recently in \cite{Cano:2023dyg} to study corrections to the extremal Kerr entropy in the context of the heterotic string and cubic gravities. For an analysis in the ungauged case see~\cite{deWit:2009de}.} 
Reproducing the black hole charges and its entropy from the near-horizon geometry is particularly useful in higher-derivative gravity, where full corrected black hole solutions are particularly difficult to obtain. Instead, the corrections to the near-horizon geometry of extremal solutions are more easily found thanks to its enhanced symmetry.% (in particular, for the case of interest to us one needs to solve algebraic rather than differential equations).

\subsection{Supersymmetric black hole and near-horizon geometry}

The most general known black hole solution to two-derivative minimal gauged supergravity in five dimensions  was given in~\cite{Chong:2005hr}. It 
 depends on four parameters $(q,m,a,b)$, and the thermodynamical quantities $\mathcal S$, $Q$, $J_1$, $J_2$ can be expressed in terms of them (see e.g.\ Section 2 of~\cite{Cassani:2022lrk} for a review of the thermodynamics). 
 Here we will restrict to the BPS solution\footnote{By BPS we denote a solution that is both extremal and supersymmetric.} with equal angular momenta, $J_1=J_2\equiv J$, that is the solution first given in~\cite{Gutowski:2004ez}. In the parameterization of~\cite{Chong:2005hr}, this depends on a single parameter $a$.
%The macroscopic entropy of this solution has been reproduced as the microscopic degeneracy of states of a class of dual SCFT in~\cite{Cabo-Bizet:2018ehj}, and for the four-derivative supergravity studied here in~\cite{Bobev:2022bjm,Cassani:2022lrk}.
%
The near-horizon geometry, also described in~\cite{Gutowski:2004ez}, is a   fibration over AdS$_2$ of a compact space with SU$(2) \times$ U$(1)$ isometry, compatible with the geometry of a three-sphere squashed by the rotation along an axis. 
Higher-derivative corrections to the near-horizon geometry preserving supersymmetry have been obtained in~\cite{Cassani:2022lrk}. The metric can be parametrized by $(t,r,\psi,\theta,\phi)$ coordinates as\footnote{$A^{\rm here}$ is $-A^{\rm there}$.}
\begin{equation}
\begin{aligned}
\text ds^2 &= v_1\left[-r^2\text dt^2 + \frac{\text dr^2}{r^2}\right] + \frac{v_2}{4}\left[\sigma_1^2 + \sigma_2^2+ v_3 \left(\sigma_3 + w\, r\text dt\right)^2\right]\,\,,\\
A &=- e\,r\,\text dt - p\left(\sigma_3 + w\,r\text dt\right)\,\,,
\end{aligned}
\label{def:ads2timess3}\end{equation}
where $\sigma$'s are the left-invariant Maurer-Cartan one-forms of SU$(2)$, given in terms of Euler angles on $S^3$ $(\psi,\theta,\phi)$ as,
\begin{equation}
\sigma_1 = \cos\psi \text d\theta + \sin\psi \sin\theta \text d\phi \,\,,\,\,\sigma_2 = -\sin\psi\text d\theta + \cos\psi \sin\theta \text d\phi \,\,,\,\,\, \sigma_3 = \text d\psi + \cos\theta \text d\phi\,\,,
\label{def:su2invariantforms}\end{equation}
and the coefficients are fixed to
\begin{equation}
\begin{aligned}
v_1 =& \frac{\chi^2}{4 g^2(1+3\chi^2)} + \alpha\,\delta v_1\,\,,\,\,\,\,\, v_2 = \frac{\chi^2}{g^2} + \alpha\, \delta v_2\,\,,\,\,\,\,\, v_3 = 1 + \frac{3}{4}\chi^2 + \alpha\,\delta v_3\,\,,\\
p =& \frac{\sqrt{3}}{4g}\chi^2 + \alpha\,\delta p \,\,,\,\,\,\,\, w = \frac{3\chi}{(1+ 3\chi^2)\sqrt{4 + 3\chi^2}} + \alpha\,\delta w\,\,,\,\,\,\,\, e = \frac{\sqrt{3}\chi}{g(1+ 3\chi^2)\sqrt{4 + 3\chi^2}} + \alpha\,\delta e\,\,,
\end{aligned}
\label{eq:nearhorizoncoefficients2d}\end{equation}
where
\begin{equation}
\begin{aligned}
\delta v_1 =&\lambda_1 \frac{18\chi^6 + 21\chi^4 + 14\chi^2 + 2}{27\chi^6 + 27\chi^4 -2}\,\,,\,\,\,\,\,\,\,\,\,\,\,\,\,\,\,\,\,\, \delta v_2 = 4\lambda_1\frac{36\chi^6 + 51\chi^4 + 10\chi^2 + 6}{9\chi^4 + 6\chi^2 - 2}\,\,,\\
\delta v_3 =& \lambda_1g^2\frac{(4 + 3\chi^2)(63\chi^4 + 78\chi^2 - 38)}{2(9\chi^4 + 6\chi^2 - 2)}\,\,,\,\,\, \delta p = \lambda_1g\frac{\sqrt{3}(4 + 3\chi^2)(27\chi^4 -6\chi^2 +10)}{4(9\chi^4 + 6\chi^2 - 2)}\,\,,\\
\delta e=&0\,\,,\,\,\,\,\, \delta w = 0\,\,.
\end{aligned}
\label{eq:nearhorizoncoefficientscorrections}\end{equation}
The solution is, therefore, given in terms of a unique parameter $\chi$, that is related to the parameter $a$ of~\cite{Chong:2005hr} by
\begin{equation}
\chi= \sqrt{\frac{2ag}{1-ag}}\,\,.
\label{eq:fromchitoa}\end{equation}

\subsection{Black hole charges from the near-horizon geometry}\label{sec:chargefromhorizon}

In Section \ref{sec:conservedchargesourtheory} we have derived the expression for the electric charge $Q$ as a  Page charge. We report here the integral we have to perform on the horizon geometry,
\begin{equation}
Q = -\frac{1}{32\pi G_{\rm eff}}\int_{\mathcal H}\sqrt{\gamma}\,n_{\mu\nu}\left[\mathcal F^{\mu\nu} +\epsilon^{\mu\nu\rho\sigma\lambda}\left( \frac{\hat c_3}{2\sqrt{3}}A_\rho F_{\lambda\sigma} + \frac{\lambda_1 \alpha}{3\sqrt{3}}\,\Omega_{\rm CS\,\rho\lambda\sigma}\right) \right]\,\,,
\label{eq:electricchargeathorizon}\end{equation}
where $\mathcal F^{\mu\nu}$ has been introduced in \eqref{eq:auxiliarytensors}.
Notice that the presence of the Lorentz-Chern-Simons term \eqref{eq:lorentzchernsimons} in this formula implies that the value of the charge is frame-dependent. The possible shifts due to a frame rotation are parametrized by an integer $k\in \mathbb Z$ as~\cite{Witten:2007kt}
\begin{equation}\label{eq:csshift}
Q \rightarrow Q + \frac{\pi\lambda_1\alpha}{4\sqrt{3}G_{\rm eff}} k\,\,.
\end{equation}
Choosing the vielbein basis
\begin{equation}
e_0 = \sqrt{v_1} r\text dt\,\,,\,\,\,\, e_1 = \sqrt{v_1}\frac{\text dr}{r} \,\,,\,\,\,\, e_2=\frac{\sqrt{v_2}}{2}\sigma_1 \,\,,\,\,\,\, e_3= \frac{\sqrt{v_2}}{2}\sigma_2\,\,,\,\,\,\, e_4 = \frac{\sqrt{v_2 v_3}}{2}\left( \sigma_3 + w\, r\text dt \right)\,\,
\end{equation}
 and evaluating \eqref{eq:electricchargeathorizon} on the near-horizon solution \eqref{def:ads2timess3}, 
\eqref{eq:nearhorizoncoefficients2d}, \eqref{eq:nearhorizoncoefficientscorrections}, we obtain
\begin{equation}
Q^* = \frac{\sqrt{3}\pi a}{2g(1-a g)^2 G_{\rm eff}}\left[1 + 4\lambda_1 \alpha g^2\frac{2\left(2 + 7 a g + 57 a^2 g^2 + 73 a^3 g^3 + 23 a^4 g^4 \right)}{3ag(11 a^2g^2 + 8 ag -1)}  \right]\,\,.
\label{eq:bpselectriccharge}\end{equation}
This can be compared with the BPS charge computed in~\cite{Cassani:2022lrk} by varying the black hole on-shell action and imposing the supersymmetric and extremal limit on the parameters. 
If we denote by $\hat Q^*$ the charge given there, we have
\begin{equation}
\hat Q^* = Q^*+ \frac{2\pi}{\sqrt{3}G_{\rm eff}}\lambda_1 \alpha\,\,,
\label{eq:comparingelectriccharges}\end{equation}
hence the two charges differ by a contribution that does not depend on the parameters of the solution (and therefore does not affect the first law of thermodynamics). We expect that this discrepancy should be due to a rotation of the vielbein causing the shift \eqref{eq:csshift}, with $k=8$.

Similarly, we can evaluate the angular momentum using the formula \eqref{eq:surfacechargeangmom} for the corresponding Page charge, that is
\begin{equation}
J = -\frac{1}{32\pi G_{\rm eff}}\int_{\mathcal H}\sqrt{\gamma}\,n_{\mu\nu} \!\left(4\nabla_\sigma \mathcal P^{\mu\nu\sigma\rho}\eta_\rho - 2 \mathcal P^{\mu\nu\sigma\rho}\nabla_\sigma\eta_\rho - \iota_\eta A\left(\mathcal F^{\mu\nu} + \frac{\hat c_3}{3\sqrt{3}}\epsilon^{\mu\nu\rho\sigma\lambda}A_\rho F_{\sigma\lambda} \right) \right),
\label{eq:angularmomentumathorizon}\end{equation}
where $\eta = 2\partial_\psi$, and $\mathcal P^{\mu\nu\rho\sigma}$ has been defined in \eqref{eq:auxiliarytensors}.
 Evaluating this formula on the near-horizon 
solution gives
\begin{equation}
J^* = \frac{a^2\left(3+ag\right)\pi}{2g\left(1-ag\right)^3 G_{\rm eff}}\left[1 + 24\lambda_1 \alpha g^2\frac{8a^4g^4 + 25 a^3g^3 + 29 a^2 g^2 + 9 ag +1}{ag(11 a^2g^2 + 8 ag -1)} \right] \,\,,
\label{eq:bpsangularmomentum}\end{equation}
which precisely matches the expression obtained from thermodynamical considerations in~\cite{Cassani:2022lrk} (note that there is no frame ambiguity in this formula).
 %The angular momentum $J^*$ in \eqref{eq:bpsangularmomentum} and the electric charge $Q^*$ in \eqref{eq:bpselectriccharge} with $k=8$ satisfy a non-linear constraint whose explicit form can be found in~\cite{Cassani:2022lrk}.

The entropy of the solution can be evaluated on the near-horizon geometry using Wald's formula \eqref{eq:Waldentropy}.
 This gives
\begin{equation}
\mathcal S^* = \frac{\pi^2}{g^3 G_{\rm eff}}\frac{ag\sqrt{ag(ag +2)}}{(1-ag)^2}\left[1 + 48\lambda_1 g^2\,\alpha\frac{2a^2g^2 + 5a g +2}{11 a^2 g^2 + 8 ag -1} \right] \,\,.
\end{equation}
The entropy can then be expressed (neglecting $\mathcal O(\alpha^2)$ terms) as a function of the BPS charges derived above as
\begin{equation}\label{eq:bpsentropy}
\mathcal S^* = \pi\sqrt{ \frac{4}{g^2} (Q^*)^2 -\frac{2\pi}{g^3 G_{\rm eff}} J^*}\,\,.
\end{equation}
This form of the microcanonical entropy is in agreement with the one found in~\cite{Cassani:2022lrk,Bobev:2022bjm}. Here we have derived it by relying just on the near-horizon solution. 

We can also check that a \emph{near-horizon version of the first law} is satisfied, where variations of the BPS entropy are related to variations of the BPS charges from the near-horizon solution,
\begin{equation}\label{nh1stlaw}
\delta\mathcal S^* = 2\pi \left(w\,\delta J^* +e\,\delta Q^*\right)\,,
\end{equation}
where $e$ and $w$ are the coefficients determined in \eqref{eq:nearhorizoncoefficients2d}.
These coefficients are non-normalizable modes of the AdS$_2$ $\times$ S$^3$ solution and correspond to potentials conjugate to the charges~\cite{Silva:2006xv,Dias:2007dj,Sen:2008yk}.
The absence of the mass in \eqref{nh1stlaw} is in line with the fact that in the BPS regime this is no more an independent quantity, as it is fixed by a linear combination of the angular momentum and electric charge.

\section{Remarks on four-dimensional supergravity}\label{sec:4d_remarks}

In this section we consider four-derivative corrections to pure $\mathcal{N}=2$ gauged supergravity in four dimensions. These have been previously studied  in \cite{Bobev:2020egg,Bobev:2021oku}, where the bosonic action was obtained by starting from conformal supergravity coupled to four-derivative invariants, gauge-fixing part of the superconformal symmetry and integrating out the auxiliary fields. The latter step consists of solving the equations of motion for the auxiliary fields in terms of the dynamical ones, and then plugging the solution back into the action.
 In order to do so, the authors of \cite{Bobev:2020egg,Bobev:2021oku}  replaced the two-derivative solution into the action, relying on the remarkable observation that solutions to the two-derivative equations of motion of the off-shell supergravity theory also solve the four-derivative equations, for any value of the higher-derivative coupling constants. However, this property alone does not allow to integrate out the auxiliary fields, since one can see that it holds true only after using the equations of motion for the metric (specifically, the trace of the Einstein equation) into one of the auxiliary field equations (specifically, the one for the field $D$ given in Eq.~(2.32) of~\cite{Bobev:2021oku}). We can still consistently integrate out the auxiliary fields using the two-derivative solution if we invoke a general argument, stating that it is legitimate to do so provided one works {\it at linear order} in the higher-derivative corrections, see e.g.~\cite{Hanaki:2006pj, Baggio:2014hua, Bobev:2021qxx, Liu:2022sew, Cassani:2022lrk}.
Working perturbatively is in fact a common way to proceed in the presence of higher-derivative couplings, as it considerably simplifies the analysis.\footnote{In practice, one is essentially forced to do this, as the auxiliary fields typically become dynamical in the higher-derivative theory; the perturbative approach allows to keep solving algebraic rather than differential equations at any order in the corrections. Moreover, this is also the natural point of view when the higher-derivative corrections are interpreted as coming from ultraviolet effects on low-energy physics, such as $\alpha'$ corrections in string theory.} 

Our observation restricts the validity of the four-derivative action given in \cite{Bobev:2020egg,Bobev:2021oku} (cf.~Eqs.~(3.1)--(3.4) of \cite{Bobev:2021oku}) to the linear order in the corrections, as ${\cal O}(\alpha^2)$ terms have been implicitly neglected when obtaining it. Taking this into account, the four-derivative action resulting from integrating out the auxiliary fields is
\begin{equation}
S \,=\, \frac{1}{16\pi G}\int\text d^4x\,e\left[\mathcal L_{2\partial} + \alpha (\lambda_1-\lambda_2) \mathcal L_{\text{Weyl}^2} + \alpha\lambda_2 \,{\cal X}_{\text{GB}}+{\cal O}(\alpha^2)\right]\,\,,
\label{eq:action4D}
\end{equation}
where
\be
\mathcal L_{2\partial}\,=\, R + 6 g^2 -\frac{1}{4}F^2
\ee
is the two-derivative Lagrangian, while the Weyl-squared Lagrangian is given by\footnote{We have added the last term which was not included in \cite{Bobev:2020egg,Bobev:2021oku}. This comes from the term $6D^2$ in eq.~(C.1) of \cite{Bobev:2021oku} and from the fact that $D=-\frac{1}{6}\left(R+12g^2\right)$.}
\be\label{eq:Weyl2}
\begin{aligned}
\mathcal L_{\text{Weyl}^2} =\,& C_{\mu\nu\rho\sigma}C^{\mu\nu\rho\sigma} -g^2 F^2 -\frac{1}{8}\left(F^2\right)^2 + \frac{1}{2}F^4+2R^\mu_{\,\,\,\nu}F^{\nu\rho}F_{\rho\mu} + \frac{1}{2}R F^2+ 2\left(\nabla_\mu F^{\mu\nu}\right)^2\\[1mm]
&+\frac{1}{6}\left(R+12g^2\right)^2\,,
\end{aligned}
\ee
where the Weyl tensor squared is conveniently expressed as
\be
C_{\mu\nu\rho\sigma}C^{\mu\nu\rho\sigma} \,=\, {\cal X}_{\text{GB}} + 2 R_{\mu\nu}R^{\mu\nu} - \frac{2}{3}R^2\,\,.
\ee
As in our previous sections, ${\cal X}_{\text{GB}}=R_{\mu\nu\rho\sigma}R^{\mu\nu\rho\sigma}-4 R_{\mu\nu}R^{\mu\nu}+R^2$ is the Gauss-Bonnet combination, $\alpha$ is a parameter with the dimension of a length square controlling the higher-derivative corrections, and $\lambda_1,\lambda_2$ are dimensionless parameters.\footnote{Our parameters are related to those used in~\cite{Bobev:2021oku} as  $c_{1,2}^{\rm there}=-\frac{\alpha}{16\pi G}\,\lambda_{1,2}^{\rm here}$.} 

We now make use of the following tensors, 
\begin{equation}\label{eq:2dEOMs}
\begin{aligned}
{\cal E}^{2\partial}_{\mu\nu}\,&=\,R_{\mu\nu}-\frac{1}{2}F_{\mu\rho}F_{\nu}{}^{\rho}+\frac{1}{8}g_{\mu\nu}F^2+3g^2 g_{\mu\nu}\,,\\[1mm]
{\cal E}^{2\partial}_{\mu}\,&=\,\nabla^{\nu}F_{\nu\mu}\,,
\end{aligned}
\end{equation}
which are set to zero by the two-derivative equations of motion of the metric and gauge field, in order to write the Weyl-squared Lagrangian \eqref{eq:Weyl2} as
\begin{equation}\label{eq:rewritingWeyl2term}
{\cal L}_{\text{Weyl}^2}\,=\,{\cal X}_{\text{GB}}+4g^2 {\cal L}_{2\partial}+2\,{\cal E}^{2\partial}_{\mu\nu}\,{\cal E}^{2\partial}{}^{\mu\nu}-\frac{1}{2}({\cal E}^{2\partial})^2+2\,{\cal E}^{2\partial}_{\mu}{\cal E}^{2\partial}{}^{\mu}\,.
\end{equation}
Using this rewriting in the action \eqref{eq:action4D}, we get
\begin{equation}\label{rewrite_4d_action}
\begin{aligned}
S \,=&\, \frac{1}{16\pi G}\int\text d^4x\,e\,\bigg\{\left[1+4\left(\lambda_1-\lambda_2\right)\alpha g^2\right]\mathcal L_{2\partial} + \lambda_2\,\alpha \,{\cal X}_{\text{GB}}\\[1mm]
&\left.+(\lambda_1-\lambda_2)\,\alpha\left[2{\cal E}^{2\partial}_{\mu\nu}\,{\cal E}^{2\partial}{}^{\mu\nu}-\frac{1}{2}({\cal E}^{2\partial})^2+2{\cal E}^{2\partial}_{\mu}{\cal E}^{2\partial}{}^{\mu}\right]+{\cal O}(\alpha^2)\right\}\,\,.
\end{aligned}
\end{equation}
Next, we note that all the terms in the second line can be neglected as their contribution to the equations of motion is effectively of order ${\cal O}(\alpha^2)$. Another way of seeing this is to notice that all these terms can be removed by making the following perturbative field redefinitions,
\begin{equation}
g_{\mu\nu}\to g_{\mu\nu}+\left(\lambda_1-\lambda_2\right)\alpha \,\left(2\,{\cal E}^{2\partial}_{\mu\nu}-\frac{1}{2}g_{\mu\nu}\,{\cal E}^{2\partial}\right)\,, \hspace{1cm} A_{\mu}\to A_{\mu}-2\left(\lambda_1-\lambda_2\right)\alpha \,{\cal E}^{2\partial}_{\mu}\, ,
\end{equation}
which become trivial on-shell. After doing so, we are left just with the first line, which we can recast as
\begin{equation}
S \,=\, \frac{1}{16\pi G_{\rm eff}}\int\text d^4 x\, e \, \mathcal L_{2\partial} 
+ \frac{\lambda_1 \alpha}{16\pi G}\int\text d^4 x\, e\, {\cal X}_{\text{GB}}
\,\,,
\label{eq:action4D4d}
\end{equation}
being $G_{\rm eff}$ the effective Newton's constant, 
\be
\frac{1}{G_{\rm eff}} = \frac{1 + 4(\lambda_1 - \lambda_2)\alpha g^2}{G}\,\,.
\ee
Since the Gauss-Bonnet term is topological in four dimensions, the equations of motion that follow from \eqref{eq:action4D4d} are thus those of the two-derivative theory, namely ${\cal E}^{2\partial}_{\mu\nu}=0$ and ${\cal E}^{2\partial}_{\mu}=0$. 

This situation is analogous to what occurs with pure gravity in four dimensions, where it is known that four-derivative corrections do not modify the solutions of the two-derivative theory, see e.g.~\cite{Endlich:2017tqa, Cano:2019ore}. It is also tantalizing to conjecture that the action of the full off-shell Poincar\'e supergravity may be rewritten as a linear combination of the two-derivative action, the Gauss-Bonnet invariant, and a sum of squares of two-derivative equations of motion, similarly to~\eqref{rewrite_4d_action}. This would make it immediately apparent that in the off-shell Poincar\'e theory the four-derivative equations are implied by the two-derivative equations.

A clear advantage of working with the field-redefined action \eqref{eq:action4D4d} is that it is straightforward to specify the boundary terms $S_{\rm GHY} + S_{\rm ct}$, where $S_{\rm GHY}$ is the generalized GHY term needed to ensure a well-defined Dirichlet variational problem and $S_{\rm ct}$ are the counterterms implementing holographic renormalization. These must be the usual boundary terms for the two-derivative action~\cite{Balasubramanian:1999re,Taylor:2000xw}, with the replacement $G \to G_{\rm eff}$, plus the generalized GHY term of \cite{Myers:1987yn} fixing the variational principle for the Gauss-Bonnet action. The latter completes the Gauss-Bonnet term to the Euler characteristic of a space with a boundary. This is a finite quantity and no new counterterms are needed beyond those renormalizing the two-derivative action. Therefore the boundary terms to be added to the bulk action are: 
\begin{equation}
\begin{aligned}
S_{\text{GHY}}&=\frac{1}{8\pi G_{\rm eff}}
 \int\diff^3x\sqrt{-h}\, {K} \\[1mm]
& -\,   \frac{\lambda_1\alpha}{4\pi G}\int%_{\partial {\cal M}}
 \diff^{3}x\,\sqrt{-h} \,\left[\frac{1}{3}K^3-K K_{ij}K^{ij}+\frac{2}{3}K_{ij}K^{jk}K_{k}{}^{i}+ 2 \Big({\cal R}_{ij}-\frac{1}{2}h_{ij}{\cal R} \Big)K^{ij}\right],
\end{aligned}
\end{equation}
and
\begin{equation}
S_{\text{ct}}= -\frac{1}{8\pi G_{\rm eff}}\int%_{\partial {\cal M}}
  \diff^3x \sqrt{-h} \, \left(2g +\frac{1}{2g}{\cal R}\right)\, ,
\end{equation}
where the notation is the same as in Section~\ref{sec:holographicrenormalization}.
These terms agree with those proposed in~\cite{Bobev:2021oku} for the original action~\eqref{eq:action4D}. In that reference, the expression for the renormalized action was reached by noting an on-shell identity relating $\int \mathcal{L}_{\rm Weyl^2}$ to a linear combination of $\int \mathcal{L}_{2\partial}$ and $\int \mathcal{X}_{\rm GB}$. Here we have shown that this identity, which is nothing but \eqref{eq:rewritingWeyl2term}, holds a priori of imposing the equations of motion, up to terms that can be eliminated with perturbative field redefinitions.

It also follows that the fixed-point theorems used in \cite{BenettiGenolini:2019jdz} to obtain a simple explicit expression for the two-derivative action evaluated on supersymmetric solutions equally apply to the four-derivative corrected action; this gives a simple proof of the conjecture made in~\cite[Sect.\:3.3]{Bobev:2021oku} and further studied in \cite{Genolini:2021urf}.
Since the Euler characteristic is just a constant, independent of the metric and the gauge field, 
 our argument also shows that the conserved charges (that can be derived by varying the renormalized action) such as the energy, the angular momentum and the electric and magnetic charges, are simply obtained from those of the two-derivative theory via the substitution $G \to G_{\rm eff}$.
This gives a direct derivation of the black hole thermodynamics in the presence of the corrections.

\section*{Acknowledgments}

We are grateful to Pablo A. Cano, Tom\'as Ort\'in, David Pere\~niguez and Dan Waldram for useful discussions. AR thanks the University of Padova and the INFN Sezione di Padova for hospitality and financial support during the final stages of this work. AR is supported by a postdoctoral fellowship associated to the MIUR-PRIN contract 2020KR4KN2, ``String Theory as a bridge between Gauge Theories and Quantum Gravity''.

\bibliographystyle{JHEP}
\bibliography{HighDer5d}

\end{document}